\documentclass[11pt]{article}
\usepackage{setspace}

\pdfoutput=1
\usepackage[utf8]{inputenc}
\usepackage[icelandic, english]{babel}
\usepackage{t1enc}
\usepackage{graphicx}
\usepackage[intoc]{nomencl}
\usepackage{enumerate,color}
\usepackage{url}
\usepackage{amsmath}
\usepackage{amsfonts}
\usepackage{thmtools}
\usepackage{amssymb}
\usepackage{amsthm}
\usepackage[nottoc]{tocbibind}
\usepackage[sf,normalsize]{subfigure}
\usepackage[format=plain,labelformat=simple,labelsep=colon]{caption}
\usepackage{placeins}
\usepackage{tabularx}
\usepackage[T1]{fontenc}
\usepackage{algorithm}
\usepackage[noend]{algpseudocode}
\usepackage{natbib}
\usepackage{booktabs}
\usepackage{rotating}
\usepackage[table,dvipsnames]{xcolor}
\usepackage{aps-bibstyle}   
\usepackage[a4paper]{geometry}
\newcommand{\fat}[1]{\boldsymbol{#1}}
\newcommand{\trp}{^\mathsf{T}}
\newcommand{\ZM}{\fat{0}}

\title{Generalization of the power-law rating curve using hydrodynamic theory and Bayesian hierarchical modeling} 

\author
{Birgir Hrafnkelsson,$^{1\ast}$ 
Helgi Sigurdarson,$^{2}$  
S{\"o}lvi R{\"o}gnvaldsson,$^{1}$ \\ 
Axel {\"O}rn Jansson,$^{1}$ 
Rafael Dan{\'i}el Vias,$^{1}$
Sigurdur Magnus Gardarsson$^{1}$ \\
\normalsize{$^{1}$University of Iceland,}\\
\normalsize{$^{2}$Isavia, Iceland}\\
}

\date{}

\begin{document}

\baselineskip12pt

\maketitle

\begin{abstract}
\noindent The power-law rating curve has been used extensively in hydraulic practice and hydrology. It is given by $Q(h)=a(h-c)^b$, where $Q$ is discharge, $h$ is water elevation, $a$, $b$ and $c$ are unknown parameters. We propose a novel extension of the power-law rating curve, referred to as the generalized  power-law rating curve. It is constructed by linking the physics of open channel flow to a model of the form $Q(h)=a(h-c)^{f(h)}$. The function $f(h)$ is referred to as the power-law exponent and it depends on the water elevation. The proposed model and the power-law model are fitted within the framework of Bayesian hierarchical models.
By exploring the properties of the proposed rating curve and its power-law exponent, we find that cross sectional shapes that are likely to be found in nature are such that the power-law exponent $f(h)$ will usually be in the interval $[1.0,2.67]$. This fact is utilized for the construction of prior densities for the model parameters. An efficient Markov chain Monte Carlo sampling scheme, that utilizes the lognormal distributional assumption at the data level and Gaussian assumption at the latent level, is proposed for the two models. The two statistical models were applied to four datasets. In the case of three datasets the generalized  power-law rating curve gave a better fit than the power-law rating curve while in the fourth case the two models fitted equally well and the generalized power-law rating curve mimicked the power-law rating curve. 
\end{abstract}
    
\newpage

\section{Introduction}\label{section1}
\pagenumbering{arabic}

Streamflow in rivers is of interest in many fields of research and applications such as \textcolor{black}{climate research \citep[e.g.,][]{meis2020}, hydroelectric power generation \citep[e.g.,][]{popescu2014}, and civil engineering design \citep[e.g.,][]{wang2015}}. Since direct methods for measuring discharge are expensive and time consuming in most cases then usually indirect methods are applied.
Indirect methods commonly involve placing a gauging station equipped with an automated hydrometer on or by a river for recording water elevation at regular time intervals. The location of a gauging station should be selected such that a channel control maintains a stable flow and the relationship between water elevation and discharge is monotonic and not prone to changes over time \citep{Mosley1993}. Estimated streamflow can then be obtained by converting time series of water elevation at the gauging station into estimated discharge by a rating curve. The rating curve is a model describing the relationship between water elevation and discharge at a given gauging station and is constructed from direct observations. 

\citet{Venetis1970} was the first to look at fitting rating curves from a statistical point of view. In \citet{Venetis1970} methods to estimate parameters and the corresponding standard errors were outlined where the rating curves were of the power-law form $Q=a(h-c)^b$, where $Q$ is discharge, $h$ is water elevation (also referred to as stage), and $a$, $b$ and $c$ are unknown parameters.  The common practice at that time was to plot discharge and water elevation measurements on a log-log paper and estimate the parameters graphically \citep[see, e.g.,][]{Herschy2009}. \citet{Clarke1999} and \citet{Clarke2000} used the classical non-linear least squares (NLS) method to derive expressions for the uncertainty of the estimated discharge to obtain uncertainties in mean annual floods and mean discharges. The methods used were essentially the same as in  \citet{Venetis1970}.
\citet{Petersen-overleir2004} proposed a model to account for heteroscedasticity in rating curve estimates. He abandoned the common practice of using a non-linear least squares on a log scale and instead proposed a model with additive errors on a real scale where both the expected discharge and standard deviation were assumed to have the power-law form.
\citet{Petersen-overleir2005} presented a method for objective segmentation in two-segmental situations as an alternative to selecting segmentation limits subjectively based on personal judgement.
\cite{Petersen-Overleir2008} fitted two segment models using global optimization to estimate parameters and bootstrap techniques to approximate uncertainty.

\citet{Moyeed2005} were the first to propose usage of the Bayesian approach for inference on rating curves. They proposed two different models for two different sets of rivers. One of the models assumed that discharge follows a Gaussian distribution where the expected discharge was given by a power-law and the other model assumed log-Gaussian distributed discharge where expected log-discharge was a linear function of water elevation.
\citet{Reitan2006} discussed shortcomings of the frequentist approach for power-law regression with a location parameter and recommended the Bayesian approach instead.
\citet{Reitan2007} proposed a statistical model which was essentially the Bayesian version of the model first described in Venetis (1970). A thorough discussion about specification of prior distributions was given as well as details of implementation and case studies.
\citet{Reitan2008} extended the Bayesian model in \citet{Reitan2007} to a multi-segment model imposing restrictions to ensure continuity of the rating curve.
\citet{Hrafnkelsson2012} proposed the assumption of smooth changes in the rating curve as an alternative to segmentation. They proposed a Bayesian model based on the models in \citet{Petersen-overleir2004}
with an added B-spline part to account for possible deviations from the power-law. 
The practice prevailing in statistical rating curve fitting, where the power function $Q=a(h-c)^b$ does not adequately describe the relationship between $Q$ and $h$, is to use segmented rating curves \citep[see, e.g.,][]{Reitan2008,Petersen-Overleir2008,Petersen-overleir2005}. 
Segmented rating curves form a flexible class of rating curves that is particularly well suited to handle shifts in the hydraulic control.

\textcolor{black}{The novelty of this paper lies in improving upon the most advanced statistical models for rating curves, i.e., those that are either based on segmentation or B-splines, by constructing a model which explicitly connects the physics of open channel flow to a generalized power-law rating curve. 
According to the formulas of Manning and Ch{\'e}zy \citep{Chow1959}, discharge is a function of the geometry of the cross-section, namely, the cross-sectional area and the wetted perimeter, and these are functions of stage. In practice, the cross-sectional area and the wetted perimeter are not available as a function of stage, however, measurements of stage are available. Given these facts and constraints, we propose a discharge rating curve of the form $Q(h)=a(h-c)^{f(h)}$, a form that can capture the Manning's formula and the Ch{\'e}zy's formula. The flexibility of this model over the power-law model comes from $f(h)$ being a function of stage while this exponent is fixed in the power-law model. Furthermore, the proposed model does not require selecting or estimating segmentation points as in the segmented rating curve models, nor selecting an upper point for the B-splines as in the B-spline rating curve models.} 

\textcolor{black}{We also propose a statistical model that can estimate the proposed discharge rating curve efficiently. In particular, by working at the logarithmic scale, a statistical model that makes use of the form $\log(Q(h))=\log(a) + f(h)\log(h-c)$ becomes feasible for inference. The functions $a(h-c)^{f(h)}$ and $f(h)$ are referred to as the generalized power-law rating curve and the power-law exponent, respectively. Through the physical formulas of Manning and Ch{\'e}zy,   
it is shown how the power-law exponent relates to the geometry of the cross-section and how the constant $a$ relates to physical parameters. This new knowledge is used to construct prior densities for the generalized power-law rating curve model. The generalized power-law rating curve and its properties have not been presented in the literature before. Same is true for the corresponding statistical model that is proposed in this paper. An efficient Bayesian computing algorithm for the proposed statistical model is presented, and the method is tested in detail on four real datasets.}

The paper is structured as follows. In Section \ref{sec:main_gplrc} the generalized power-law rating curve \textcolor{black}{is introduced, its connection to the physics of flow in open channels is derived and its mathematical properties are explored}. The four real datasets on \textcolor{black}{pairs of discharge and stage are introduced in Section \ref{chDATA}. In Section \ref{chModelsandInference} a statistical model based on the generalized power-law rating curve is proposed for this type of data and its inference scheme is introduced}. 
The proposed statistical model is applied to the four datasets and the results are presented in Section \ref{rebbs} and conclusions are drawn in Section \ref{chConclusion}.   

\section{The generalized power-law rating curve}
\label{sec:main_gplrc}

In this section we introduce the generalized power-law rating curve. First\textcolor{black}{, in Section \ref{sec:meanvelo}}, physical models for mean velocity \textcolor{black}{and discharge} in open channels are reviewed. \textcolor{black}{In Section \ref{sec:afef}} the form of the generalized power-law rating curve is \textcolor{black}{formally} proposed and its relationship to the underlying physics and the cross-section geometry is derived. Finally, \textcolor{black}{in Section \ref{sec:properties},} the properties of the generalized power-law rating curve are explored\textcolor{black}{, and we show how knowledge about these properties can be used to construct prior densities for this rating curve model}.

\subsection{Models for mean velocity \textcolor{black}{and discharge} in open channels}
\label{sec:meanvelo}

\begin{figure}[b!]\centering
\begin{minipage}{0.4\linewidth}
\includegraphics[width=\linewidth]{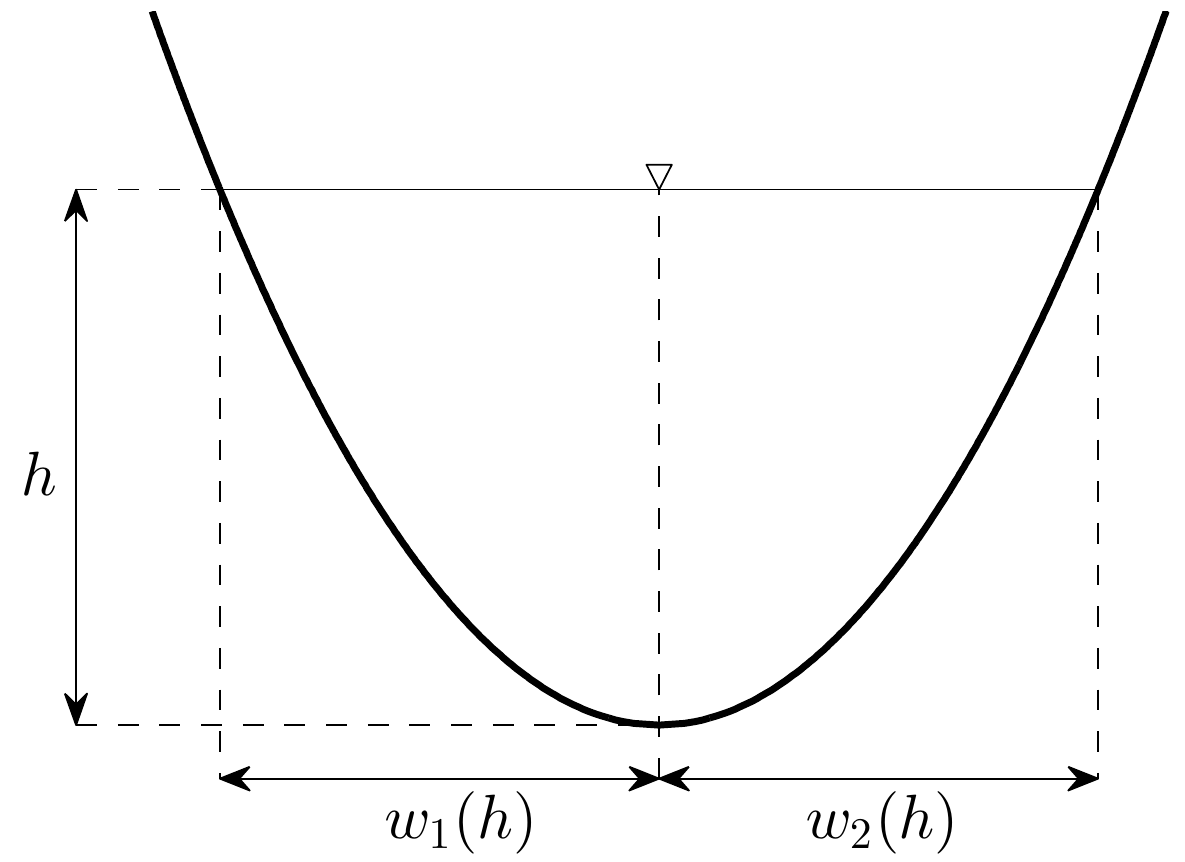}\end{minipage}
\caption{\textcolor{black}{The cross-sectional area $A$ at water depth equal to $h$, $A(h)$, is the area defined by the curve and the upper horizontal line. The wetted perimeter $P$ at water depth equal to $h$, $P(h)$, is the length of the curve below the upper horizontal line. The total width of a cross section at water depth $h$, $w(h)$, is the sum of the horizontal lengths $w_1(h)$ and $w_2(h)$.}}
\label{fig:cross_sec_fig}
\end{figure}

\textcolor{black}{In this subsection we go through the formulas of Manning and Ch{\'e}zy for mean velocity in open channels \citep{Chow1959} and show how they depend on the cross-sectional area, $A$, and the wetted perimeter, $P$, defined as the circumference of the cross section excluding the free surface, \textcolor{black}{see Figure \ref{fig:cross_sec_fig}}. Note that both $A$ and $P$ depend on stage, $h$. By multiplying these formulas with the cross-sectional area, formulas for discharge are obtained. These discharge formulas are the product of physical constants and a geometry factor that changes with stage. In practice, estimation of discharge rating curves for open channels in nature is based on paired observations of discharge and stage, but not on observations of cross-sectional area and wetted perimeter since these are usually not collected. Thus, estimation of discharge rating curves cannot be based directly on the formulas of Manning and Ch{\'e}zy, and rating curves based on stage only, such as the generalized power-law rating curve, are needed. However, to understand how the properties of the proposed generalized power-law rating curve relate to the physics of flow in open channels, it is essential to have the general form of the physical discharge formulas since they include the physical parameters and the geometry.} 

Open channel flow has a free surface subject to atmospheric pressure as opposed to closed conduit flow, for example a pipe flow, where the flow is pressurized. The velocity of the flow is not uniform over a given cross section; however, the mean velocity through the cross section is of interest in practice as it can be used to compute the discharge through a cross section, given the cross-sectional area. The discharge, and therefore the mean velocity, is governed by the balance between the gravitational force and a force due to frictional resistance \citep{Chow1959}. 

The Ch{\'e}zy formula, developed by the French engineer Antoine Ch{\'e}zy, was derived from hydrodynamic theory \citep{Chow1959}. It gives the mean velocity, $\overline{v}$, at a cross section in a uniform, gravity-driven, fully developed, turbulent flow in open channels. 
Uniform flow refers to flow in a channel where the cross section, the friction and the slope remain constant in the flow direction.  This is of course rarely the case in natural channels and is therefore an assumption which does not strictly hold; however, it is often quite satisfactory. Ch{\'e}zy's formula is given by 
$$
\overline{v}=CR^{1/2}S^{1/2}
$$
where $C$ is Ch{\'e}zy's constant, representing the frictional resistance, $S$ is the slope of the channel and $R$ is the hydraulic radius. The hydraulic radius is defined as the ratio between the cross-sectional area, $A$, and the wetted perimeter, $P$, \textcolor{black}{i.e., $R=A/P$.}
Assuming SI units, $R$ and $P$ are in meters, $\overline{v}$ in m s$^{-1}$, $A$ is in m$^2$, $S$ is unit free and $C$ is in m$^{1/2}$ s$^{-1}$.

The Irish engineer Robert Manning presented an empirical formula for the mean velocity of a uniform, gravity-driven, fully developed, turbulent flow in rough open channels based on experimental data \citep{Chow1959}. Manning's formula is given by
$$
\overline{v}=\frac{1}{n}R^{2/3}S^{1/2}
$$
(assuming SI units) where $n$ is the Manning roughness coefficient (s m$^{-1/3}$). Values of Manning roughness coefficient for various types of channel surfaces and their roughness can be found in handbooks \citep[see, e.g.,][]{Shen1993}. \citet{Gioia2001} derived Manning's empirical formula theoretically using the phenomenological theory of turbulence. 

Ch{\'e}zy's and Manning's formulas are linked through $C=\frac{1}{n}R^{1/6}$, implying that Ch{\'e}zy's $C$ is a function of $R^{1/6}$. Other authors have suggested that in natural channels $n$ is a function of $R$ to some power. For example, in the case of gravel-bed rivers where the slope exceeds $0.002$, \textcolor{black}{the equation $n = 0.32 S^{0.38} R^{-0.16}$ was suggested by} \citet{Jarrett1984} when SI units are used. \textcolor{black}{Another formula of this type is $n=(f)^{1/2}(8g)^{-1/2}R^{1/6}$} \citep{Herschy2009} where $g$ is the Earth's gravitational acceleration, $f$ is the Darcy--Weisbach friction factor. This formula is the result of equating the Manning and Darcy--Weisbach equations.

Assuming either Ch{\'e}zy's or Manning's formula where $C$ or $n$ can depend on the hydraulic radius to some power, the mean velocity through a cross section can then be written as
\begin{equation}\label{eq:gen1}
\overline{v}=k R^x
\end{equation}
where $k$ and $x$ are constants independent of the water elevation. Furthermore, discharge can then be written as
\begin{equation}\label{eq:gen1dis}
Q=\overline{v}A=kR^xA=k\frac{A^{x+1}}{P^x}.
\end{equation}
Note that in this form, neither formula is assumed over the other nor is it assumed that Ch{\'e}zy's $C$ or Manning's $n$ depend on some power of $R$; rather (\ref{eq:gen1}) is a generalized form of Ch{\'e}zy's and Manning's formulas that takes into account the possibility that $C$ or $n$ may be functions of $R$ to some power. Ch{\'e}zy's and Manning's formulas with constant $C$ and $n$ are special cases of  (\ref{eq:gen1}); while assuming $C$ and $n$ are a function of $R$ to some power, the two formulas then coincide in (\ref{eq:gen1}). This formula is similar to what has been called the \textcolor{black}{generalized friction law, $\overline{v}=K_1 R^xS^{1/2}$}
\citep{petersen2006modelling}. \citet{Chow1959} \textcolor{black}{also noted that most uniform-flow formulas are of the general form $\overline{v}=K_2 R^xS^{y}$.
In the equations above $K_1$ and $K_2$ are constants independent of the water elevation.}

\textcolor{black}{Table \ref{hydroquanties} in Appendix \ref{app0} contains a list of the hydrodynamic quantities and parameters found in this subsection along with their units.}

\subsection{Generalization of the power-law rating curve}
\label{sec:afef}

\textcolor{black}{In this subsection we formally propose the generalized power-law rating curve.
It is a generalization of the power-law rating curve of the form
\begin{equation}\label{eq:gpl}Q(h)=a(h-c)^{f(h)},
\end{equation}
where $a$ and $c$ are constants and $f(h)$ is the power-law exponent. The motivation for the form of the generalized power-law rating curve in (\ref{eq:gpl}) is given below, and it is demonstrated how physics of open channels flow as presented by (\ref{eq:gen1dis}) enter into (\ref{eq:gpl}), i.e., what is the physical interpretation of $a$ and how does the geometry in (\ref{eq:gen1dis}) affect $f(h)$.}   

Although the power-law formula is empirical, it stems from theory. In particular, in the case of a v-shaped cross section, both the Ch{\'e}zy formula and the Manning formula yield a power-law rating curve with $b=2+x$.
For other shapes, there is no direct link between the power-law rating curve and the formulas of Ch{\'e}zy and Manning. 
\textcolor{black}{The fact that the power-law formula is exact for uniform flow in the case of a v-shaped cross section} indicates that an extension of the power-law form might be a sensible form for rating curves in general. 
The logarithmic transformation of the power-law form
gives a form that is linear in terms of the parameters
$\log(a)$ and $b$, that is,
$\log Q(h)=\log(a)+b\log(h-c)$, which is convenient for 
statistical inference. This model can be extended by
allowing either one of $\log(a)$ and $b$ to vary with $h$, 
or both of them.
By modeling $\log(a)$ and $b$ as a function of $h$ with a linear statistical model of some sort, the statistical inference will be easier for that model compared to a model that assumes nonlinear
forms for $\log(a)$ and $b$. 
It is \textcolor{black}{shown} below that by allowing only $b$ to vary with water elevation, a flexible form of a rating curve can be developed which captures the physical nature of discharge in open channels.

To incorporate the physics of open channel flow into the generalized power-law rating curve in (\ref{eq:gpl}), the general formula for discharge in uniform flow given in (\ref{eq:gen1dis}) and the rating curve in (\ref{eq:gpl}) are equated with $c=0$ for simplicity. Thus, in this subsection \textcolor{black}{and in Section \ref{sec:properties},} $h$ will represent the water depth. So, without any loss of generality,
\begin{equation}
Q(h)=ah^{f(h)}=k\frac{A(h)^{x+1}}{P(h)^x}.
\nonumber
\end{equation}
Solving for the power-law exponent $f(h)$ and $a$ gives Result 1 below.
Here $A(h)$ and $P(h)$ denote the cross-sectional area and the wetted perimeter, respectively, as a function of water depth $h$, and they are defined as
\begin{equation}
A(h)=\int^h_0w_1(\eta)d\eta + \int^h_0w_2(\eta)d\eta, 
\nonumber
\end{equation}
\begin{equation}\label{eq:gplc_temp}
P(h)=\int_0^h \sqrt{1+\left\{w_1'(\eta)\right\}^2}d\eta+\int_0^h \sqrt{1+\left\{w_2'(\eta)\right\}^2}d\eta
\end{equation}
where $w_1(h)$ and $w_2(h)$ are two lengths that together that make up the width of the cross section at water depth $h$, see Figure \ref{fig:cross_sec_fig}.
The terms 
$w_1'(h)$ and $w_2'(h)$ are the first derivatives of $w_1(h)$ and $w_2(h)$ with respect to $h$. It is assumed that $w_1(h)$ and $w_2(h)$ are continuous, and that $w'_1(h)$ and $w'_2(h)$ are piecewise continuous to ensure that the integrals for $A(h)$ and $P(h)$ exist and are continuous. Furthermore, it is assumed that the cross section is such that it forms a single area for all values of the water depth, i.e., there cannot be two or more disjoint areas for any value of the water depth. This means that $w_1(h)$ and $w_2(h)$ are always positive and can only take one value for each water depth $h$.  \\

\noindent
\textbf{Result 1} The power-law exponent $f(h)$ \textcolor{black}{in (\ref{eq:gpl})} is given by
\begin{equation}\label{eq:gplc}
f(h)=\frac{ (x+1)\log
\left\{\displaystyle\frac{A(h)}{A(1)}\right\}
-x\log
\left\{\displaystyle\frac{P(h)}{P(1)}\right\}}{\log(h)}
\end{equation} 
for $h>0$ and $h \neq 1$\textcolor{black}{. The} constant $a$ in (\ref{eq:gpl}) is given by
$$a=k\frac{A(1)^{x+1}}{P(1)^x}=Q(1).$$

A proof of Result 1 is given in Appendix \ref{app1}. 
Assuming that the model in (\ref{eq:gen1dis}) gives an accurate description of discharge in a uniform gravity driven fully developed turbulent flow in open channels with constant friction and constant slope in the flow direction, the model in (\ref{eq:gpl}) is simply another way to rewrite the model in (\ref{eq:gen1dis}) \textcolor{black}{given that $w_1(h)$ and $w_2(h)$ are restricted to being positive and taking only one value for each water depth $h$.} So, under these constrictions, the model in (\ref{eq:gpl}) is as flexible as the model in (\ref{eq:gen1dis}). That means the model given by (\ref{eq:gpl}) can be used to model 
\textcolor{black}{any regular or irregular geometry in the cross section of open channels that falls under the constrictions}. 

Note that $f(h)$ is affected by the geometry of the cross section and $x$ but not by the parameter $k$. Since $k$ is a function of the friction ($C$ or $n$) and the slope ($S$), $f(h)$ is not affected by the friction nor the slope.
The constant $a$ is equal to $Q(1)$\textcolor{black}{, i.e., discharge when the depth is equal to $1$ m}, and that gives the simplest interpre

The proposed model in (\ref{eq:gpl}) can be used as a basis
for a statistical model of the form 
\begin{equation}
\log(Q_i)=\log(a)+f(h_i)\log(h_i-c) + \epsilon_i
\label{eq:statmodel}
\end{equation}
where $(h_i,Q_i)$ are the $i$-th water elevation/discharge observation and $\epsilon_i$ is the corresponding error term.
There are several ways to specify a model for $f(h)$ within this statistical model. The assumptions given for the model in (\ref{eq:gpl}) which involve $w_k(h)$ being continuous and $w'_k(h)$ being piecewise continuous, $k=1,2$, can be used as a reference. These assumptions lead to $f(h)$ being continuous and $f'(h)$ being piecewise continuous, since $f(h)$ is a function of $P(h)$ and $A(h)$ and the first derivative of $P(h)$ is a function of the first derivative of $w_k(h)$, $k=1,2$, while the first derivative of $A(h)$ is a function of $w_k(h)$, $k=1,2$. So a finite number of jumps in $f'(h)$ could be allowed in a given interval over $h$. 
Statistical models with more restrictive constraints on $f(h)$ than above may be more feasible for statistical inference, for example;
(i) $f(h)$ and $f'(h)$ are continuous; (ii) $f(h)$, $f'(h)$ and $f''(h)$ are continuous. As previously noted, a linear statistical model for $f(h)$ is desired, so,
models that are linear in the statistical parameters, and fulfil one of the three restrictions presented above, are candidates for $f(h)$ in the statistical model given by (\ref{eq:statmodel}).

\subsection{Properties of the generalized power-law rating curve}
\label{sec:properties}

In this \textcolor{black}{subsection} the properties of the generalized power-law rating curve are explored through the power-law exponent $f(h)$.
Important properties of the power-law exponent $f(h)$ are its limits as $h$ approaches zero from above, one and infinity. 
These limits are given in Result 2. \\

\noindent
\textbf{Result 2}
Assume that $w''_1(h)$ and $w''_2(h)$ are continuous. The values of $f(h)$ at $h=0$ and $h=1$ are defined as the limit of $f(h)$ as $h$ approaches $0$ from above and as $h$ approaches $1$, respectively. That is,
\begin{equation}\label{eq:gplb2}
f(0) = \lim_{h\rightarrow 0^{+}} f(h) = 1 + (x+1)\lim_{h\rightarrow 0^{+}}\frac{hA''(h)}{A'(h)} 
- x \lim_{h\rightarrow 0^{+}} \frac{h P''(h)}{P'(h)}
\end{equation}
and 
\begin{equation}\label{eq:gplb3}
f(1) = \lim_{h\rightarrow 1} f(h) = (x+1)\frac{A'(1)}{A(1)} - x \frac{P'(1)}{P(1)},
\end{equation}
furthermore, the limit of $f(h)$ as $h$ approaches infinity is given by
\begin{equation}\label{eq:gplb4}
\lim_{h\rightarrow \infty} f(h) = 1 + (x+1)\lim_{h\rightarrow \infty}\frac{hA''(h)}{A'(h)} 
- x \lim_{h\rightarrow \infty} \frac{h P''(h)}{P'(h)}
\end{equation}
where 
$$
A'(h) = w_1(h)+w_2(h),  \quad A''(h) = w_1'(h)+w_2'(h), 
$$
$$
P'(h) = \sqrt{1+ \left\{ w_1'(h) \right\}^2} + \sqrt{1+ \left\{ w_2'(h) \right\}^2}
$$
and
$$
P''(h) = \frac{w_1'(h)w_1''(h)}{\sqrt{1+ \left\{ w_1'(h) \right\}^2}} + \frac{w_2'(h)w_2''(h)}{\sqrt{1+ \left\{ w_2'(h) \right\}^2}}.
$$

The proof for Result 2 is shown in Appendix \ref{app2}. 
For further insight into the generalized power-law rating curve, its exponent function $f(h)$ is investigated for simple cross section shapes 
assuming a uniform flow.
The simple cross section shapes considered here are symmetric ($w_1(h)=w_2(h)$) and the cross section width, $w_{\alpha}(h)$, is a power function of water depth, 
$$w_{\alpha}(h)=w_1(h)+w_2(h)=\phi_{\alpha}h^{\alpha},\qquad \alpha \geq 0,$$
where $\alpha$ is a cross-sectional shape parameter and $\phi_{\alpha}$ is a positive constant which defines the width of the cross section at $h=1$.
A few general results are derived for these cross section shapes below. The cross-sectional area corresponding to $w_{\alpha}(h)$ is given by 
$$
A_{\alpha}(h)=\int^h_0w_{\alpha}(\eta)d\eta=\int^h_0 \phi_\alpha\eta^\alpha d\eta=\frac{\phi_\alpha}{\alpha+1} h^{\alpha+1}
$$
and the wetted perimeter is given by
$$
P_{\alpha}(h)=2\displaystyle\int_0^h\textstyle\sqrt{1+4^{-1}\alpha^2 \phi_{\alpha}^2 \eta^{2(\alpha-1)}}d\eta
$$
as $w_1(h)=w_2(h)=0.5 w_{\alpha}(h)$.
The power-law exponent corresponding to a symmetric cross section with the width $w_{\alpha}(h)$
is denoted by $f_{\alpha}(h)$. Results for $f_{\alpha}(h)$ are given in Result 3. \\

\noindent
\textbf{Result 3} The form of $f_{\alpha}(h)$ according to (\ref{eq:gplc}) is given by 
\begin{equation}\label{eq:gold}
f_{\alpha}(h)=(x+1)(\alpha+1)-x
\frac{\{\log P_{\alpha}(h) - \log P_{\alpha}(1) \} }
{\log(h)}.
\end{equation} 
The limit of $f_{\alpha}(h)$ as $h$ approaches zero from above is
\begin{equation}\label{eq:gpld}
\lim_{h\rightarrow 0^+}f_{\alpha}(h)=f_{\alpha}(0)=
 \left\{ \begin{array}{ll}
         \alpha + x+1 & \mbox{if $0 \leq  \alpha \leq 1$,}\\
         \alpha + 1 + \alpha x & \mbox{if $\alpha > 1$}.\end{array} \right. 
\end{equation}
The limit of $f_{\alpha}(h)$ as $h$ approaches one is
\begin{equation}\label{eq:gple}
\lim_{h\rightarrow 1}f_{\alpha}(h)=f_{\alpha}(1)=
(x+1)(\alpha+1)-x\frac{P'_{\alpha}(1)}{P_{\alpha}(1)},
\end{equation}
where 
$P'_{\alpha}(1) = 2\sqrt{1+4^{-1}\alpha^2 \phi_{\alpha}^2 }$.
The limit of $f_{\alpha}(h)$ as $h$ approaches infinity is
\begin{equation}\label{eq:gplf}
\lim_{h\rightarrow \infty}f_{\alpha}(h)=
 \left\{ \begin{array}{ll}
         \alpha + 1 + \alpha x & \mbox{if $0 \leq  \alpha \leq 1$,}\\
         \alpha + x+1 & \mbox{if $\alpha > 1$}.\end{array} \right. 
\end{equation}

\noindent
The proof for Result 3 is shown in Appendix  \ref{app3}. 

\begin{figure}[b!]
\centering
\begin{minipage}{0.4\linewidth}
\includegraphics[width=\linewidth]{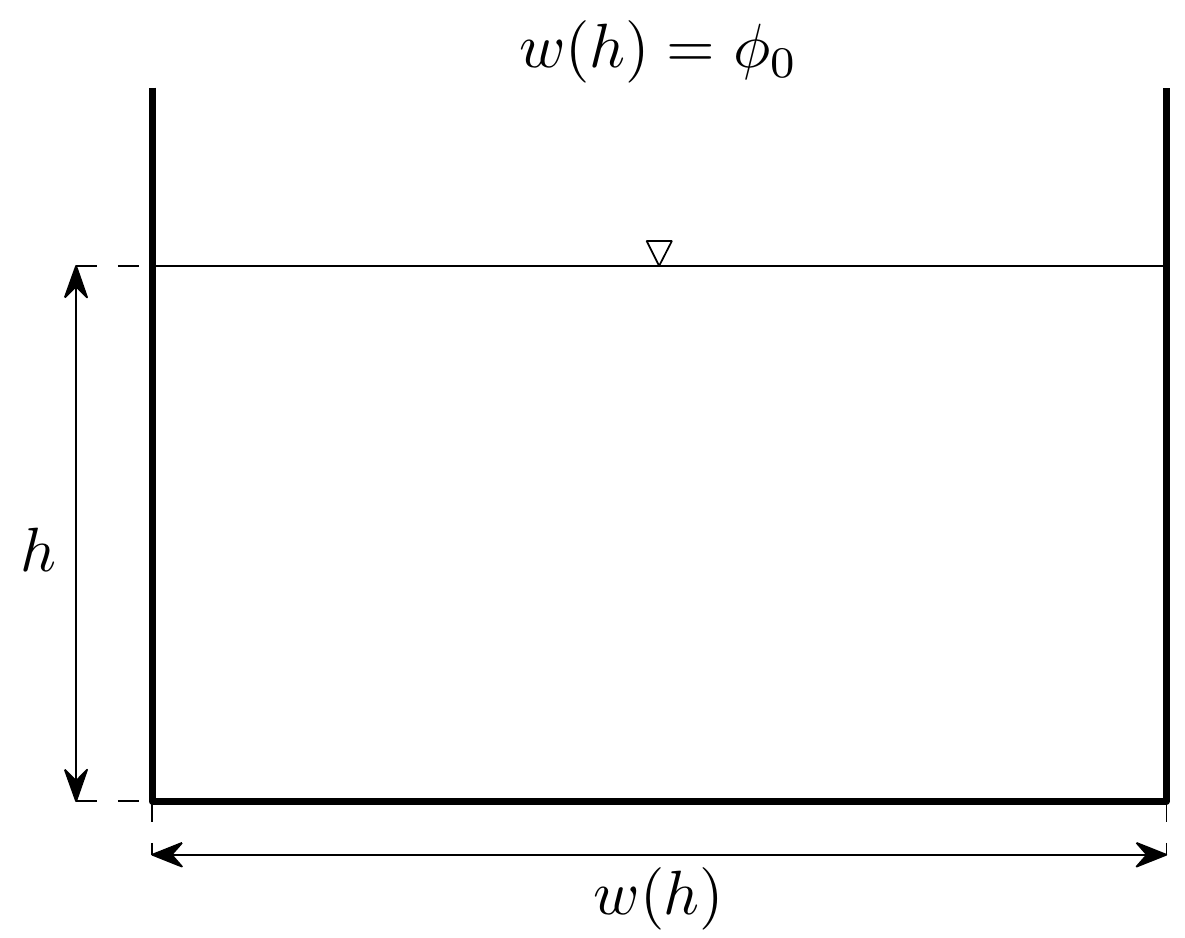}
\end{minipage}
\begin{minipage}{0.4\linewidth}
\includegraphics[width=\linewidth]{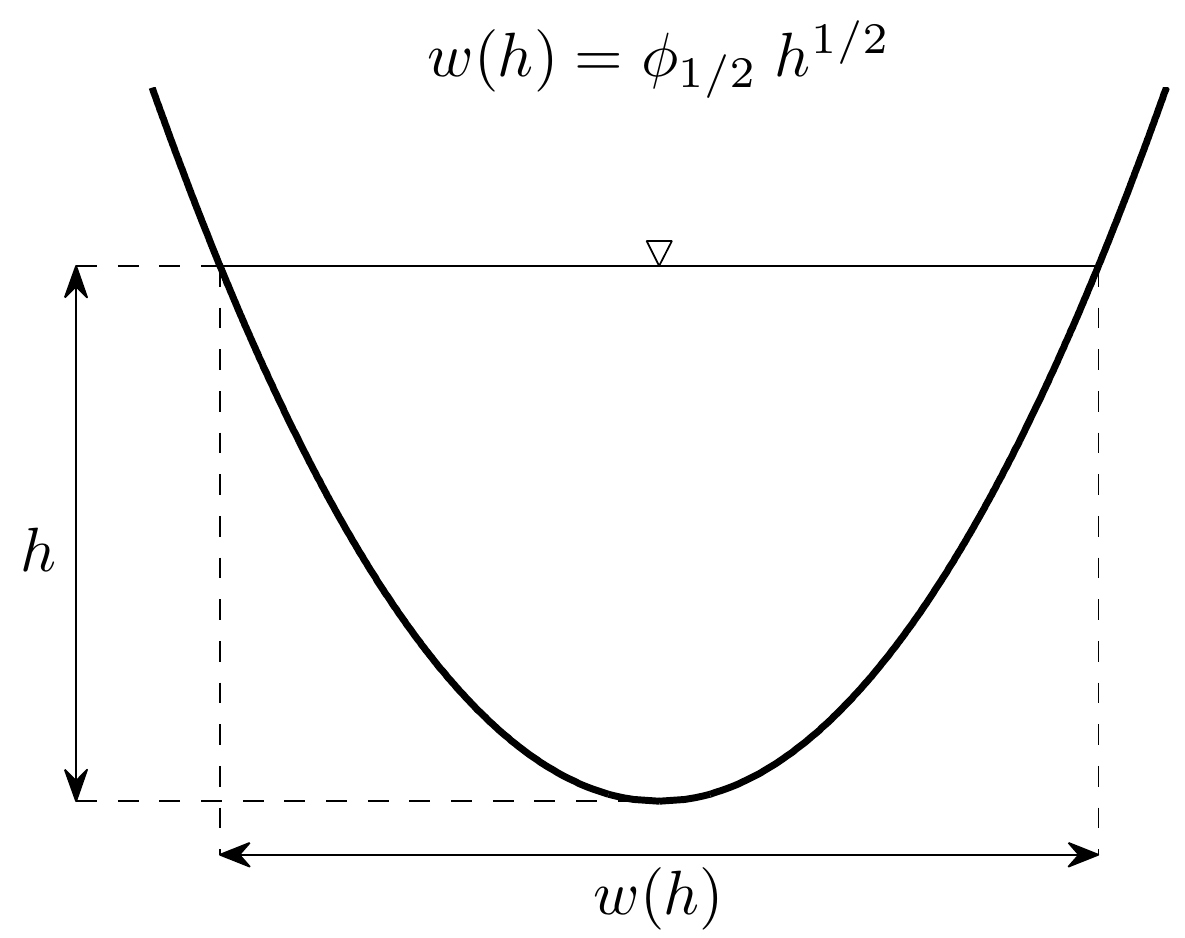}
\end{minipage}\vspace{5mm} \\
\begin{minipage}{0.4\linewidth}
\includegraphics[width=\linewidth]{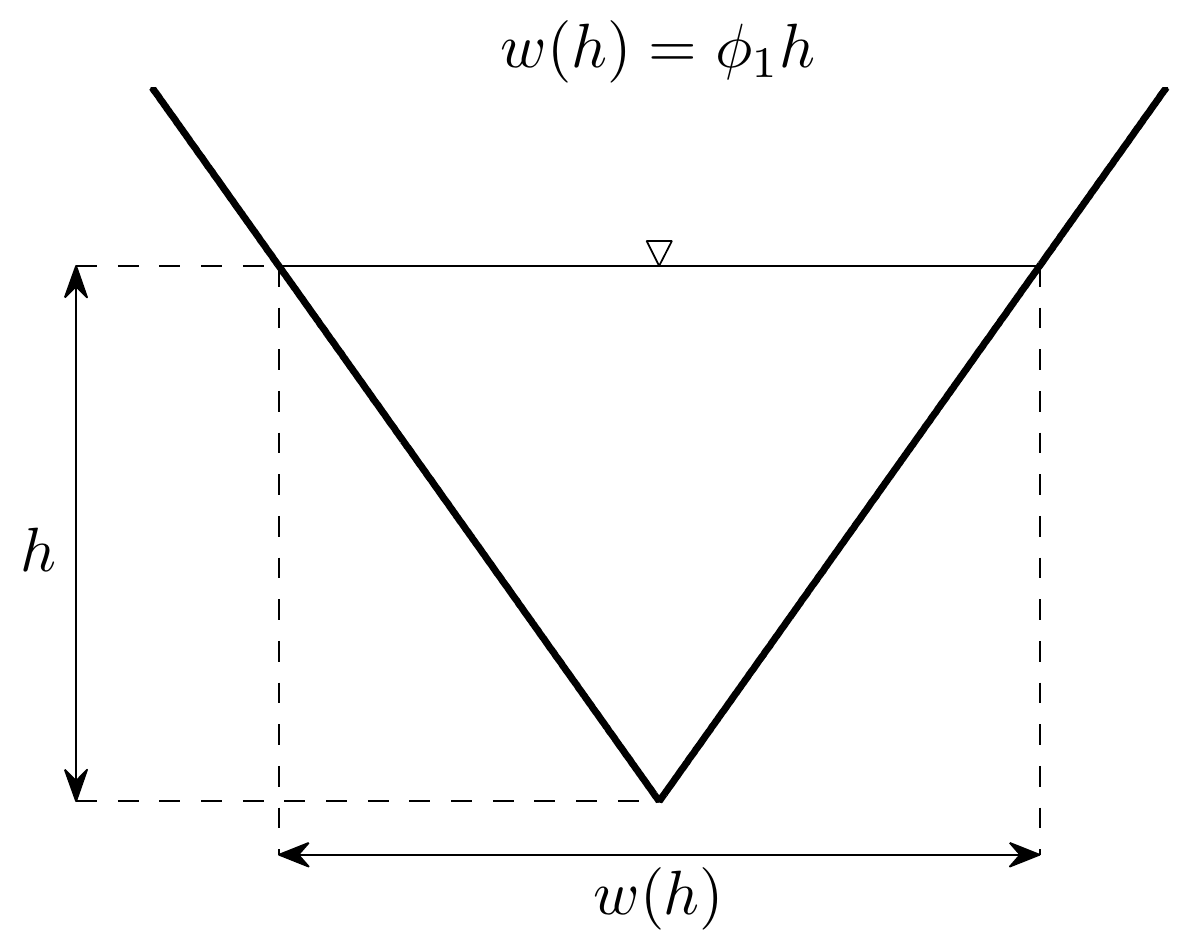}
\end{minipage}
\begin{minipage}{0.4\linewidth}
\includegraphics[width=\linewidth]{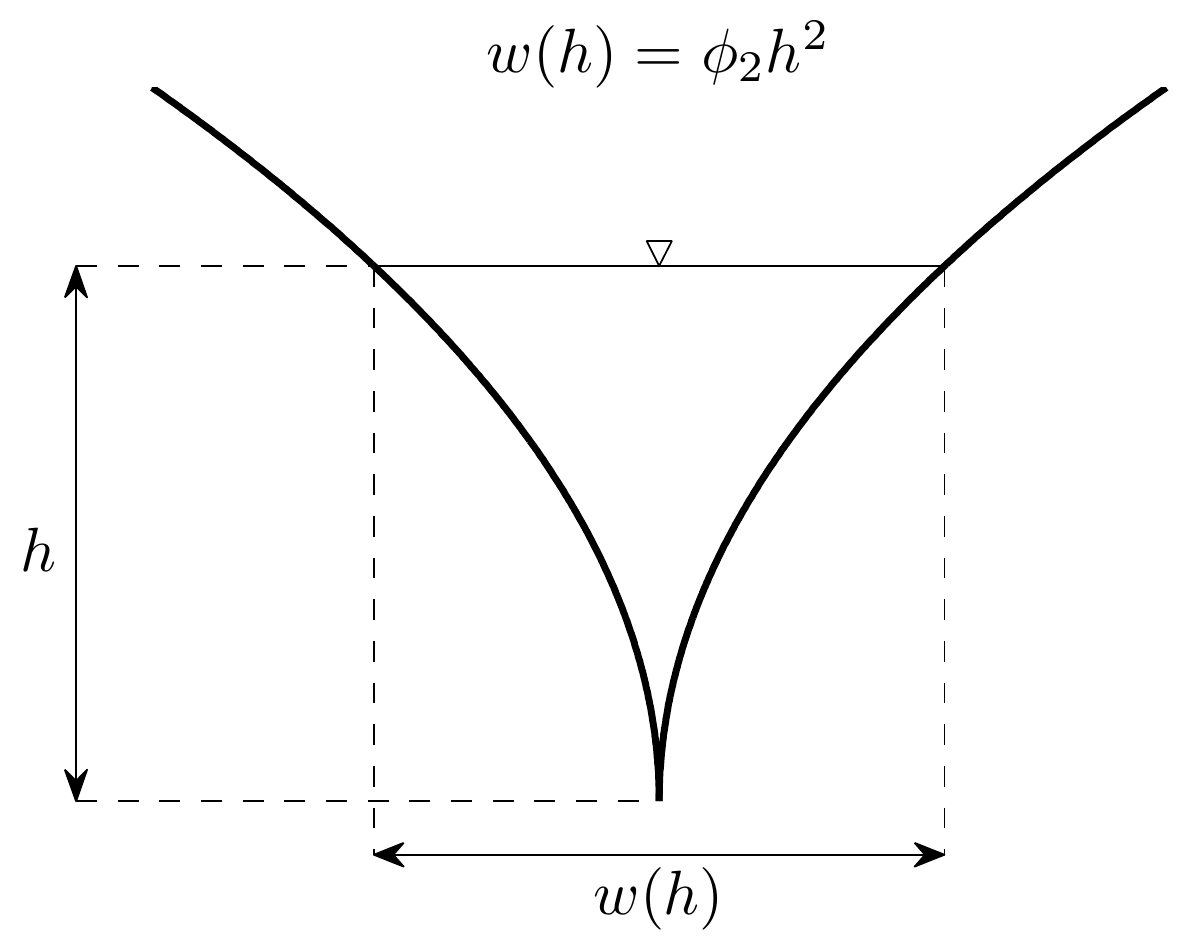}
\end{minipage}
\caption{Four simple cross sections; rectangular ($\alpha=0$) (top left panel); parabolic ($\alpha=1/2$) (top right panel); triangular ($\alpha=1$) (bottom left panel); inverse parabolic (bottom right panel) ($\alpha=2$).}
\label{fig:4sh}
\end{figure}

\begin{figure}[h]
\centering
\includegraphics[width=0.85\linewidth]{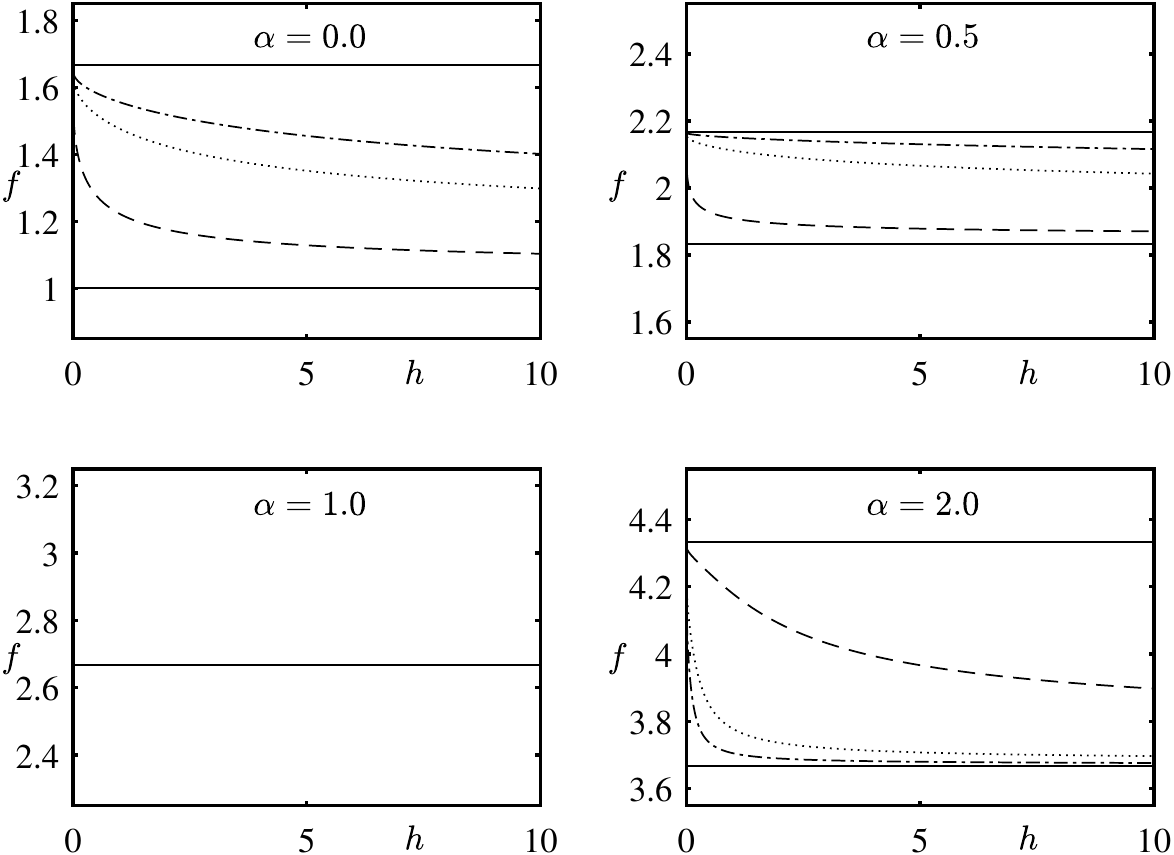}
\caption{$f_{\alpha}(h)$ for the four simple shaped cross sections with $\alpha \in \{0,1/2,1,2 \}$ for three values of $\phi_\alpha$;  
$\phi_\alpha = 1$ (dashed); $\phi_\alpha=5$ (dotted); $\phi_\alpha=10$ (dashdot);  assuming $x=2/3$. The straight lines show the limits as $h\rightarrow0^{+}$ (upper line) and as $h\rightarrow\infty$ (lower line).}\label{fig:breg}
\end{figure}

Below are results based on Result 3 derived for four cross section shapes corresponding to $\alpha\in\left\{0,\frac{1}{2},1,2\right\}$. The four shapes are shown in Figure \ref{fig:4sh}. Note that $\alpha=0$ corresponds to the rectangular cross section as $w_0(h)=\phi_0 h^0= \phi_0$. The values $\alpha=\frac{1}{2}$, $\alpha=1$ and $\alpha=2$ correspond to parabolic, triangular and inverse parabolic cross sections, respectively. The cross-sectional areas for these cross section shapes are given by 
$$
A_0(h) = \phi_0 h, \quad A_{1/2}(h) = \frac{2}{3}\phi_{1/2}h^{3/2},
$$
$$
A_1(h) = \frac{1}{2}\phi_1 h^2, \quad A_{2}(h) = \frac{1}{3}\phi_{2}h^{3}.
$$
In the case of the four cross sections with $\alpha\in\left\{0,\frac{1}{2},1,2\right\}$, the wetted perimeter is given by 
$$
P_0(h) = \phi_0 + 2 h,
$$
$$
P_{1/2}(h) = 2\sqrt{h}\sqrt{h + (\phi_{1/2}/4)^2} 
+ 2 (\phi_{1/2}/4)^2 \log\left(\frac{\sqrt{h}+\sqrt{h + (\phi_{1/2}/4)^2}}{\phi_{1/2}/4} \right),
$$
$$
P_1(h) = 2h\sqrt{1+4^{-1} \phi_{1}^2 } , \quad 
P_{2}(h) = h\sqrt{1+\phi_2^2h^2}+\phi_2^{-1}\log\left(\phi_2 h+\sqrt{1+\phi_2^2h^2}\right).
$$
In the case of the triangular cross section, $\alpha=1$, the exponent $f_{\alpha}(h)$ is
constant with respect to $h$, that is,
$$
f_1(h)=2+x,
$$
which corresponds to $b=2+x$ in the power-law model as noted in the beginning of Section \ref{sec:afef}.
For the rectangular cross section then $\alpha=0$ and the exponent $f_{\alpha}(h)$ is
$$
f_0(h)=(x+1)-x\frac{\{\log(\phi_0+2h) - \log(\phi_0+2)\}}
{\log(h)},
$$
and its limits as $h$ approaches zero and infinity are $(x+1)$ and $1$, respectively,
and the limit of $f_0(h)$ at $h=1$ is
$$
f_0(1)=(x+1) - 2x/(2+\phi_0).
$$
The exponent $f_{\alpha}(h)$ and its limit at $h=1$ for $\alpha=1/2$ and $\alpha=2$ are found by evaluating (\ref{eq:gple}) using $P_{1/2}(h)$, $P_{1/2}(1)$, $P'_{1/2}(1)$ for $\alpha=1/2$ and
$P_{2}(h)$, $P_{2}(1)$, $P'_{2}(1)$ for $\alpha=2$.
The limits of $f_{1/2}(h)$ as $h$ approaches zero and infinity are $3/2+x$ and $3/2+x/2$, respectively.
The limits of $f_{2}(h)$ as $h$ approaches zero and infinity are $3+2x$ and $3+x$, respectively.
Figure \ref{fig:breg} shows $f(h)$ for the four shapes for varying values of $\phi_\alpha$ assuming $x=2/3$. The function $f_{\alpha}(h)$ is bounded by its limits at zero and infinity as defined by (\ref{eq:gpld}) and (\ref{eq:gplf}).

\begin{figure}[h]
\centering
\includegraphics[width=0.85\linewidth]{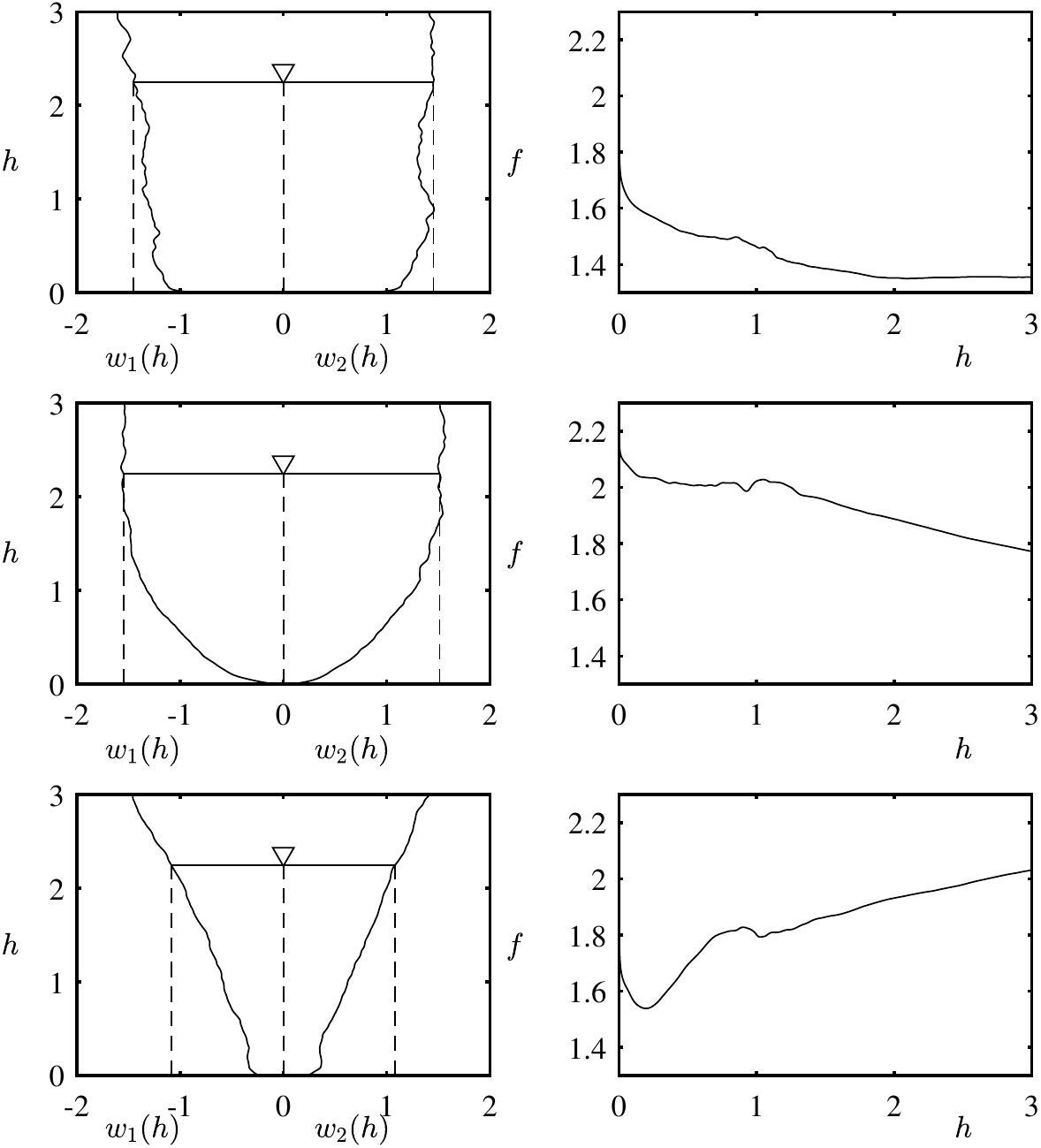}
\caption{The left panel shows three cross sections with irregular geometry, the water depth is on the y-axis and the width (in terms of $w_1(h)$ and $w_2(h)$) is on the x-axis. The right panel shows their corresponding power-law exponents on the y-axis as function of the water depth. It is assumed that $x=2/3$.}
\label{fig:irreg_cross_sec_fig}
\end{figure}

The results in (\ref{eq:gpld}) and (\ref{eq:gplf}) are important in terms of understanding the behavior of the power-law exponent, $f_\alpha(h)$, for the cross sections with shapes given by the width $w_\alpha(h)$ that has the simple mathematical representation $\phi_\alpha h^\alpha$. These shapes can be used to approximate shapes found in natural settings, in particular, those corresponding to $\alpha \in (0,1)$, i.e., from rectangular shape, through parabolic shape to triangular shape, see Figure \ref{fig:4sh}. 
For example, in the case of the parabolic cross section ($\alpha=1/2$), the function $f(h)$ is bounded between $1.5 + 0.5x$ and  $1.5+x$, or $1.83$ and  $2.17$ if $x=2/3$. The simple cross sections corresponding to $\alpha \in (0,1)$ suggest that cross sections found in nature are likely to have generalized power-law rating curves that are such that their power-law exponent $f(h)$ takes values between $1.0$ (the lower  bound of the rectangular shape) and $2.67$ (the upper  bound of the triangular shape).  


Figure \ref{fig:irreg_cross_sec_fig} shows three examples of geometry that mimic what can be found in nature, and the corresponding power-law exponent, $f(h)$. Note that in Figure \ref{fig:irreg_cross_sec_fig} it is assumed that $x=2/3$. The model in (\ref{eq:gpl}) can handle the geometry in Figure \ref{fig:irreg_cross_sec_fig} and the corresponding power-law exponent can be computed. 
The left top panel of Figure \ref{fig:irreg_cross_sec_fig} shows a cross section that is close to a rectangular shape with small variation as $h$ increases. This results in a power-law exponent that is close to the one stemming from an exact rectangular shape as seen in Figure \ref{fig:breg}. The small scale variation from the rectangular shape in the cross section have little effect on the power-law exponent. 
In the middle panel of Figure \ref{fig:irreg_cross_sec_fig}, the shape of the cross section is close to a parabolic shape for $h$ between $0$ m and $2$ m, and for $h$ greater than $2$ m the cross section is close to being vertical. The corresponding power-law exponent takes values between $2.0$ and $2.2$ when $h$ is small but gradually decreases as $h$ increases. This can be explained by the transition from a parabolic shape to vertical shape as in the rectangular shape. The pure parabolic and rectangular shapes give a power-law exponent with values equal to $2.13$ and $1.67$ for small values of $h$, respectively, and decreases towards the values $1.83$ and $1.0$ as $h$ becomes larger, respectively (see Figure \ref{fig:breg}). So, the power-law exponent of the cross section in the middle panel decreases gradually from a value close to $2.13$ and when $h$ is equal to $3$ m, the power-law exponent is down to a value below $1.8$ m.
The cross section shown in the bottom panel is close to a v-shape but with a flat bottom. So, for small values of $h$ the shape is more like a rectangular shape while the v-shape becomes more apparent as $h$ increases. The corresponding power-law exponent is thus taking a value close to $1.67$ when $h=0$ m and decreases for small values of $h$, however, as $h$ becomes larger, it starts to increase and at $h=3$ m it is greater than $2.0$. This is not surprising since the power-law exponent of the pure v-shape is equal to $2.67$ for all $h$, and the power-law exponent in the bottom panel would approach that value if the v-shape would also hold for larger water depth.

The results above are important for the selection of prior densities for parameters associated with $f(h)$ when modeling open channel flow in natural settings within a Bayesian statistical framework, namely, whatever parameterization is used for $f(h)$, it should be such that the selected prior densities of the parameters place $f(h)$ in the interval $[1.0, 2.67]$ with high probability. The model for $f(h)$ and the prior densities of the parameters associated with $f(h)$ will be introduced in Section \ref{chModels}.

\section{Data}\label{chDATA}

Four datasets were considered for a detailed analysis. They consist of paired observations of discharge and stage. Each dataset belongs to a specific observational site in Iceland. The data were collected by the Icelandic Meteorological Office (IMO) from rivers with quite diverse conditions at different locations in Iceland. 
The rivers are the Nordura River that runs through the Borgarfjordur region in central west Iceland (number of pairs $n=35$); the Skjalfandafljot River, which has a source in the northwest of the Vatnajokull Icecap from where it flows north into Skjalfandi Bay in central part of north Iceland ($n=56$); the Jokulsa a Fjollum River located in the northeast of Iceland, its source being the Vatnajokull Icecap ($n=76$); and the fourth river is the Jokulsa a Dal River in eastern Iceland which now contains a reservoir for hydroelectric power generation, with its source being 
the Bruarjokull Icecap ($n=86$). The Nordura River is a spring water river with direct runoff components, and the other three rivers are glacial rivers. These rivers were the subject of a previous study described in \cite{Hrafnkelsson2012}. 

\begin{figure}[hbt!]\centering
\begin{minipage}{0.9\linewidth}
\includegraphics[width=\linewidth]{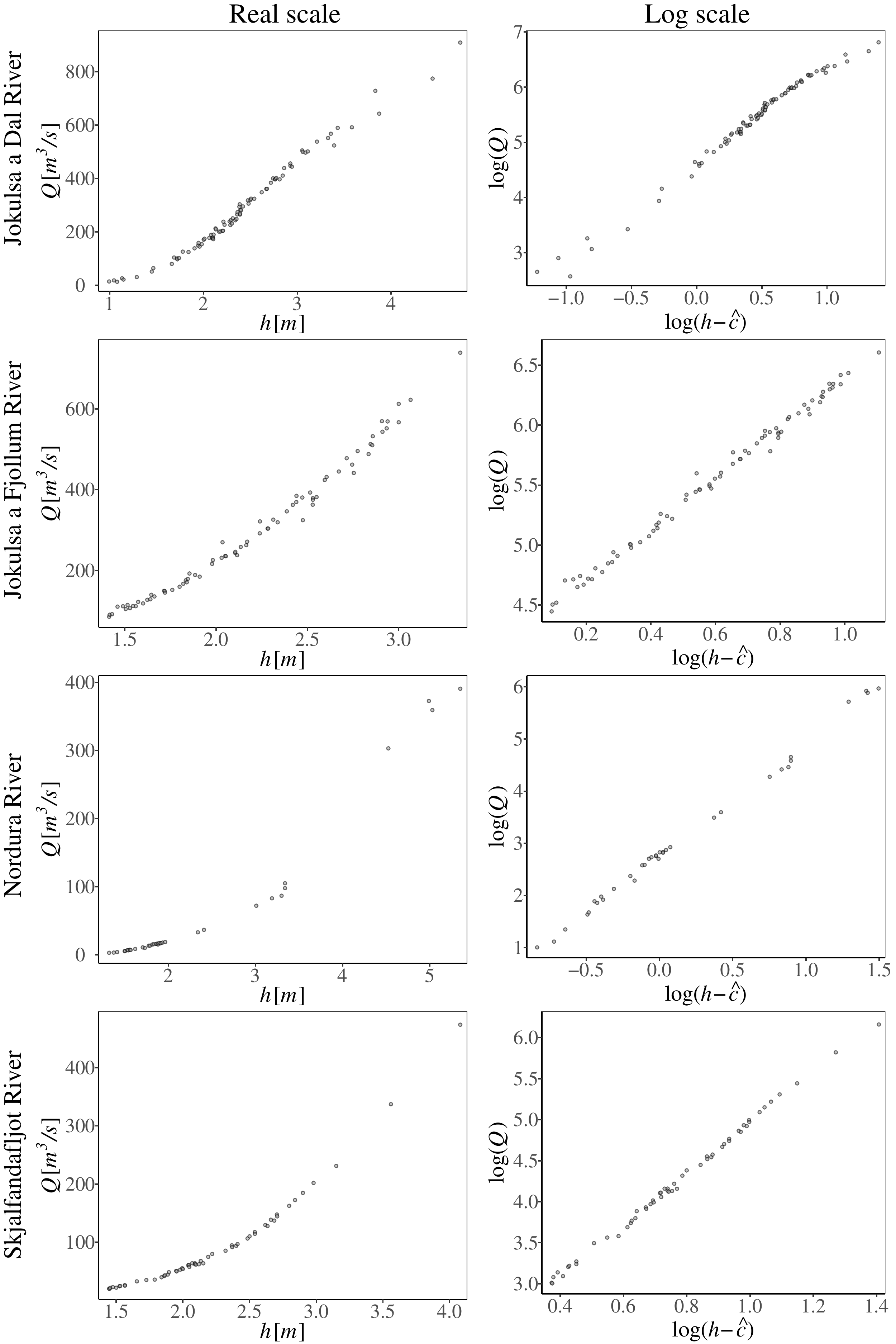}
\end{minipage}
\caption{\textcolor{black}{Discharge versus stage for each of the four rivers (left panel). The logarithmic transformation of discharge versus the logarithmic transformation of the difference between stage, $h$, and $\hat{c}$, an estimate of the stage where discharge is zero (right panel).}}
\label{fig:data_sec_fig}
\end{figure}

\textcolor{black}{Figure \ref{fig:data_sec_fig} shows discharge versus stage in the left panel for the four rivers, and the right panel shows the logarithmic transformation of discharge versus the logarithmic transformation of the difference between stage and $\hat{c}$ where $\hat{c}$ is an estimate of the stage where discharge is zero, i.e., the posterior median of $c$ under the power-law model. The plots in the right panel are such that when the power-law rating curve is an adequate model then the data cluster around a straight line, and an estimate of its slope is an estimate of $b$ in the power-law rating curve. The data from the Jokulsa a Fjollum River can be model adequately well with a straight line while that is not the case for the data from the Jokulsa a Dal River. The data from the other two rivers appear to deviate from a straight line. Analysis of these four dataset in the Results section reveals which of them can be described adequately well with the power-law rating curve and which require the generalized power-law rating curve.}

\section{Statistical modeling and inference}\label{chModelsandInference}

In this section we propose a statistical model based the generalized power-law rating curve along with an efficient Bayesian inference scheme for the model.
This statistical model, referred to as Model 1, will be compared to a statistical model based on the power-law rating curve, referred to as Model 0. 

\subsection{Bayesian models for discharge rating curves}\label{chModels}

The proposed statistical model for discharge observation, Model 1, assumes that its median is given by the generalized power-law rating curve, that is ,
$$Q(h)=a(h-c)^{f(h)},$$
where, as before, the power-law exponent $f(h)$ is a function of stage, $h$, the parameter $c$ is the stage at which the discharge is zero while the parameter $a$ is a scaling parameter that can be interpreted as the discharge when the corrected stage, $h-c$, is equal to $1.0$ m. The power-law exponent is parameterized as a sum of a constant $b$ and deviations $\beta(h)$, that is,
\begin{align*}f(h)=b+\beta(h) .
\end{align*}
The $i$-th discharge observation \textcolor{black}{$\tilde{Q}_i$}, conditional on its corresponding stage, $h_i$, is modeled \textcolor{black}{as} a lognormal variable, 
\begin{equation}\label{eq:datadens}
\textcolor{black}{\tilde{Q}_i} \sim \textrm{LN}(\log(a)+\{b+\beta(h_i)\}\log(h_i-c),\sigma_\epsilon^2(h_i)), \quad i = 1,...,n,
\end{equation}
\textcolor{black}{where $\sigma^2_{\epsilon}(h_i)$ is the variance of the $i$-th measurement error at the logarithmic scale and $n$ is the number of observations. This variance is allowed to vary with stage since a preliminary analysis of several datasets of paired discharge and stage observations revealed that some datasets are such that the variance of the residuals varies with stage while other datasets are such that it is reasonable to assume it is a constant. Examples of these two cases can be seen in the results section.} The reduced version of Model 1, i.e., Model 0, is such that $\beta(h_i)=0$ for all $i$, \textcolor{black}{and thus, the median takes} the form of the traditional power-law rating curve. \textcolor{black}{The variance of Model 0 varies with stage}.

Figure \ref{fig:breg} provides the values of the power-law exponent, $f_{\alpha}(h)$, in the generalized power-law rating curve for the symmetric cross sections defined by the width $w_{\alpha}(h) \propto h^{\alpha}$ while Figure \ref{fig:irreg_cross_sec_fig} demonstrates what the exponent $f(h)$ could look like in natural settings. 
It is reasonable to assume that the forms given by $w_{\alpha}(h)$ are close to the forms found in nature, in particular those with $0 \leq \alpha \leq 1$, where $\alpha=0$, $\alpha=1/2$ and $\alpha=1$, correspond to the rectangular, parabolic and triangular shapes, respectively. Thus, based on our new knowledge about $f(h)$ in Section \ref{sec:properties}, we know that its values will most likely lie in the interval $[1.0,2.67]$. 
In Section \ref{sec:afef} it is argued that a sensible model for $f(h)$ assumes that $f(h)$ is either once differentiable or twice differentiable.
We opt for the latter choice. The latter model is smoother than the former model and thus better suited for smoothing the noise found in the observations. The form of $f(h)$ will not be known before hand, thus, a flexible model such as a two times mean square differentiable Gaussian process is proposed as a prior for $f(h)$. Further details on the prior densities of the model parameters will be given below.   

The error variance, $\sigma^2_\epsilon(h)$, of the log-discharge data, under both Model 0 and Model 1, is modeled as an exponential of a B-spline curve, that is, a linear combination of B-spline basis functions, $B_k$, $k=1,...,K$, \citep{wassermann2006all} that are defined over the range of stage observations or
\begin{equation}\label{eq:sigmah}
\sigma_\epsilon^2(h)=\exp\left\{\sum_{k=1}^K \eta_k B_k(h)\right\}=\prod_{k=1}^K \exp\left(\eta_k B_k(h)\right)
\end{equation} 
where $\eta_1$, ..., $\eta_K$ are unknown parameters and $K$ is the number of basis functions. The basis functions are defined on the interval $[h_{\min},h_{\max}]$ where $h_{\min}$ and $h_{\max}$ are the smallest and largest stage observations in the paired dataset, respectively. Furthermore, $\sigma_\epsilon^2(h_{\min})=\exp(\eta_1)$ and $\sigma_\epsilon^2(h_{\max})=\exp(\eta_K)$.
\textcolor{black}{The interior knots are equally spaced on the interval $[h_{\min},h_{\max}]$ while the additional knots are set equal to the end points of the interval.
This model can capture the case of a constant variance with respect to stage within $[h_{\min},h_{\max}]$ since the right-hand side of (\ref{eq:sigmah}) is constant when $\eta_1=\eta_2 =... =\eta_K$.}

\textcolor{black}{To facilitate calculations of discharge predictions corresponding to a new pair of observed stage and discharge for any value of the stage,
the error variance is defined outside of the interval $[h_{\min},h_{\max}]$ as $\sigma_\epsilon^2(h)=\exp(\eta_1)$ for $h<h_{\min}$ and as $\sigma_\epsilon^2(h)=\exp(\eta_K)$ for $h>h_{\max}$.
This is a simple model and its purpose is to provide a prediction interval for discharge that can be used as a reference when $h$ is outside of $[h_{\min},h_{\max}]$. 
Note that the primary interest lies in the rating curve itself, i.e., the median of the model in (\ref{eq:datadens}), which is used to transform time series of stage to discharge. The error variance on the interval $[h_{\min},h_{\max}]$ affects the posterior variance of the rating curve but the specification of the error variance outside this interval does not affect its posterior variance.} 

The parameter $a$ represents the discharge (m$^3$/s) when the depth of the river is $1.0$ m. We opt for a weakly informative prior density for this quantity since we can rely on the data to inform about its value. It is very likely that $a$ is in the interval $[10^{-2},10^5]$. 
The logarithmic transformation of $a$ is used in the inference scheme.
A Gaussian prior density with mean $\mu_a=3.0$ and
standard deviation $\sigma_a=3.0$ is selected for $\log(a)$ since it represents the interval above reasonably well.

Based on arguments above and in Section \ref{sec:main_gplrc} it is reasonable to assume a priori that for a given $h$ there is a high probability of $f(h)=b+\beta(h)$ being in the interval $[1.0,2.67]$, say $0.95$. We opt for fixing $b$ at the central value of this interval, i.e., set $b=1.835$, and count on $\beta(h)$ to capture the variability in $f(h)$ after fixing $b$. This is achieved by lining up the prior densities of $\beta(h)$ and its associated parameters such that they support reasonable shapes of $b+\beta(h)$, e.g., the once seen in Figures  \ref{fig:breg} and \ref{fig:irreg_cross_sec_fig}. Under a Gaussian assumption and using $0.95$ as a reference probability of being in the interval $[1.0, 2.67]$, the standard deviation of $\beta(h)$ is $0.426$. However, since the standard deviation of $\beta(h)$ is an unknown parameter then the value $0.426$ will be used as a reference value, see below. Under the reduced model, Model 0, it is assumed that the function $\beta(h)$ is zero, and under this model the parameter $b$ is assigned a Gaussian prior density with mean $1.835$ and standard deviation $0.426$.

In accordance with our assumption of a twice differentiable $f(h)$, we propose modeling the function $\beta(h)$ with a mean zero Gaussian process that is governed by a Mat\'ern covariance function with marginal standard deviation $\sigma_\beta$, range parameter $\phi_\beta$ and smoothness parameter $\nu$. 
The amplitude of the process is controlled by $\sigma_\beta$, 
$\phi_\beta$ governs how fast the spatial correlation of the process decays with increasing distance $d$ and $\nu$ controls the smoothness of the process. In general, if $2<\nu \leq 3$ then the Mat{\'e}rn process is two times mean-square differentiable. Here the value 
$\nu=5/2$ is selected as that correlation function has a relatively simple form.
The prior density of the vector $\fat{\beta}=(\beta_1,...,\beta_n)\trp$ where $\beta_i=\beta(h_i)$ is Gaussian with mean zero and covariance matrix $\Sigma_\beta$,
i.e., $\fat{\beta} \sim \textrm{N}(\ZM,\Sigma_\beta)$, where the $(i,j)$-th element of $\Sigma_\beta$ is
\begin{equation}\label{eq:matcov}
\{\Sigma_\beta\}_{i,j}=\textrm{Cov}(\beta(h_i),\beta(h_j))=\sigma^2_\beta \left(1 + \frac{\sqrt{5}d_{i,j}}{\phi_\beta} +\frac{5 d_{i,j}}{3\phi^2_\beta}  \right) \exp\left(-\frac{\sqrt{5}d_{i,j}}{\phi_\beta}  \right)
\end{equation} 
where $d_{i,j}=|h_i-h_j|$ is the distance between stages $h_i$ and $h_j$. The prior densities for $\sigma_\beta$ and $\phi_\beta$ are described below.

The parameter $c$ is assigned a prior density that is such that the quantity $h_{\min}-c$ has an exponential density with rate parameter $\lambda_1$ where  $h_{\min}$ is either the smallest stage in the paired dataset $(h_1,Q_1)$, ..., $(h_n,Q_n)$ or the smallest stage found in other data sources for \textcolor{black}{the river under study}.
Thus, the prior density for $c$ is set after seeing data on stage only while the posterior density of $c$ is determined based on the paired dataset.
Here, after exploring other datasets from Iceland, $\lambda_1$ is selected such that the probability of $h_{\min}-c$ being greater than $1.5$ m is $0.05$ which corresponds to $\lambda_1=2$. In the case of datasets from other areas, the value of $\lambda_1$ should be reconsidered. The parameter $c$ is transformed to $\psi_1=\log(h_{\min}-c)$. The prior density for $\psi_1$ is 
$$
\pi(\psi_1)=\lambda_1 \exp(-\lambda_1 \exp(\psi_1) + \psi_1), \quad \psi_1 \in \mathbb{R}.
$$

The prior density for $(\sigma_\beta,\phi_\beta)$ is based on the Penalised Complexity (PC) prior 
density for the Mat{\'e}rn parameters derived in \citet{fuglstad2019constructing}.
The PC approach for the selection of prior densities is based on setting up a base model,
i.e., a simpler model that is reasonable to shrink towards \citep{simpson2017penalising}. Here, in the case of $\beta(h)$ and
its parameters, using the PC prior density of \citet{fuglstad2019constructing}, means that the base model is the one with 
the range equal to infinity and the marginal standard deviation equal to zero. So, if the data suggest that $\beta(h)$
is constant in the observed interval of stages then the PC prior density supports that, i.e., it supports that one over the range takes values arbitrary close to zero. 
Or, if the data suggest that the marginal standard deviation is zero or relatively small then the PC prior density supports these
scenarios as well. Thus, the joint prior density of $(\fat{\beta},\sigma_\beta,\phi_\beta)$ allows realizations of $\fat{\beta}$ that are close to being constant or smoothly varying in the observed interval of stages, and such that $f(h)$ is most likely in the interval $[1.0, 2.67]$

The form of the prior density for $(\sigma_\beta,\phi_\beta)$ is given by
$$
\pi(\sigma_\beta,\phi_\beta)=
\lambda_2 \exp(-\lambda_2 \sigma_\beta) \lambda_3 (2\phi_\beta)^{-3/2} \exp(-\lambda_3 (2 \phi_\beta)^{-1/2}),
$$
where the parameter $\rho$ in \citet{fuglstad2019constructing} is such that $\rho=2\phi_\beta$. 
The values of the parameters $\lambda_2$ and $\lambda_3$ are found by specifying
the reference value $\sigma_{\beta,0}$ for $\sigma_\beta$, and the reference value $\phi_{\beta,0}$ for $\phi_\beta$, and assuming that the probability of $\sigma_{\beta}$ 
being  above $\sigma_{\beta,0}$ is $\alpha_2$, and that the probability of $\phi_{\beta}$ 
being below $\phi_{\beta,0}$ is $\alpha_3$. Here the following values are selected
$$
\sigma_{\beta,0}=0.426, \quad \alpha_2=0.10, \quad \phi_{\beta,0}=1.5, \quad \alpha_3=0.10,
$$ 
where $\phi_{\beta,0}$ is in meters. Selecting $\sigma_{\beta,0}=0.426$ is related to the arguments above, namely, conditional on $\sigma_{\beta}=0.426$, 
the probability of $b+\beta(h)$ being outside the interval $[1.0,2.67]$ is $0.05$. If $\sigma_{\beta}$ is greater than $0.426$ then this probability increases, so, setting $\alpha_2$ equal to $0.10$ equates putting small prior probability on that scenario.     
 
Furthermore, having the range parameter above $1.5$ m with prior probability $0.90$ ensures slow changes in $\beta(h)$ within a relatively short interval, e.g., of length $0.5$ m.   
Then $\lambda_2$ and $\lambda_3$ are
$$
\lambda_2= -\log(\alpha_2)(\sigma_{\beta,0})^{-1}=5.405, \quad \lambda_3=-\log(\alpha_1)(2\phi_{\beta,0})^{1/2}=3.988.
$$
The parameters $\sigma_\beta$ and $\phi_\beta$ are transformed to $\psi_2=\log(\sigma_\beta)$ and $\psi_3=\log(\phi_\beta)$. The prior density for $(\psi_2,\psi_3)$ is
\begin{equation}
\pi(\psi_2,\psi_3)=
\lambda_2\exp(-\lambda_2 \exp(\psi_2) + \psi_2) \\
\lambda_3 2^{-3/2} \exp(-1.5\psi_3) \exp(-\lambda_3 (2)^{-1/2}\exp(-0.5\psi_3) + \psi_3),
\nonumber
\end{equation}
where $\psi_2,\psi_3 \in \mathbb{R}$. \textcolor{black}{Since $\pi(\psi_2,\psi_3)$ factorizes it can written as $\pi(\psi_2,\psi_3)=\pi(\psi_2)\pi(\psi_3)$.}

The parameter $\eta_1$ is directly linked to the standard deviation of
the error term when $h=h_{\min}$ through $\sigma_{\epsilon}(h_{\min})=\exp(0.5\eta_1)$, $h_{\min}$ being the smallest stage in the paired data . 
\citet{simpson2017penalising} found that the PC prior density for the standard deviation in a Gaussian density, when the base model has a standard deviation equal to zero, is an exponential density. We believe a priori that $\sigma_{\epsilon}(h_{\min})$ can be arbitrary close to zero, thus, this PC prior density is appropriate for this parameter. Furthermore, by exploring other datasets, we find it reasonable to set the prior density for $\sigma_{\epsilon}(h_{\min})$ such that it is likely to be below the value $0.08$. Thus, we calibrate the exponential prior density for $\sigma_{\epsilon}(h_{\min})$ such that
the probability of exceeding the value $0.08$ is $0.10$, giving the rate is $\lambda_{5}=28.78$. This is equivalent of $\eta_1$ having the prior density
$$
\pi(\eta_1)=\frac{1}{2}\lambda_{5} \exp\left(-\lambda_{5} \exp(0.5\eta_1) + 0.5\eta_1  \right), \quad \eta_1 \in \mathbb{R}. 
$$  
The prior density of $\fat{\eta}_{-1}=(\eta_2,...,\eta_K)$, conditional on $\eta_1$ and the standard deviation parameter $\sigma_{\eta}$, is specified in terms of Gaussian densities, that is,
$$
\pi(\fat{\eta}_{-1}|\eta_{1},\sigma_{\eta})=\prod_{k=2}^{K}\textrm{N}(\eta_k|\eta_{k-1},\sigma^2_{\eta}),
$$
which is a random walk prior.
This prior density of $\fat{\eta}_{-1}$ can be presented as
$$
\pi(\fat{\eta}_{-1}|\eta_{1},\sigma_{\eta})=(2\pi\sigma^2_\eta)^{-(K-1)/2}\exp\left(-\frac{1}{2\sigma^2_\eta}\fat{\eta}\trp R_{\eta}\fat{\eta}\right),
$$
where $\fat{\eta}=(\eta_1,...,\eta_K)$.
The prior density of $\fat{\eta}$, conditional on $\sigma_{\eta}$, is $\pi(\fat{\eta}|\sigma_{\eta})=\pi(\eta_{1})\pi(\fat{\eta}_{-1}|\eta_{1},\sigma_{\eta})$.
Here $K$ is set equal to $6$.
In the case of $K=6$ then
\begin{equation}
R_{\eta} = \begin{bmatrix}    1 & -1 &  0 &  0 &  0 &  0  \\ 
                             -1 &  2 & -1 &  0 &  0 &  0 \\
                              0 & -1 &  2 & -1 &  0 &  0  \\ 
                              0 &  0 & -1 &  2 & -1 &  0 \\ 
                              0 &  0 &  0 & -1 &  2 & -1 \\                                                          
                              0 &  0 &  0 &  0 & -1 &  1 \end{bmatrix},
\nonumber
\end{equation}
see \citet[Section 3.3.1, p. 95]{rue2005gaussian}.

Again, motivated by the PC approach of \citet{simpson2017penalising} for the selection of prior densities,
we assign an exponential prior
density to the standard deviation parameter $\sigma_{\eta}$. 
This is in line with our modeling approach as we believe that the standard  deviation $\sigma_{\epsilon}(h)$ can be a constant in some cases 
and that corresponds to $\sigma_{\eta}$ being equal to zero.
This prior density is selected such that the probability of $\sigma_{\eta}$ 
exceeding the value $0.267$ is $0.10$ which corresponds to
a rate parameter $\lambda_4=8.62$.
This is motivated by the fact that when the value of $\sigma_{\eta}$ is $0.267$
then for $K=6$ the value of $\exp(0.5\eta_K)$, i.e., the standard deviation at $h=h_{\max}$ ($h_{\max}$ being the largest stage in the paired data),  
can become more than two times larger
than $\exp(0.5\eta_1)$,  
with probability $0.01$ and it can become less than half of the size of $\exp(0.5\eta_1)$ with probability $0.01$.
The parameter $\sigma_\eta$ is transformed to $\psi_4=\log(\sigma_\eta)$. The prior density for $\psi_4$ is 
$$
\pi(\psi_4)=\lambda_4 \exp(-\lambda_4 \exp(\psi_4) + \psi_4), \quad \psi_4 \in \mathbb{R}.
$$  

With the aim of improving the sampling scheme for the posterior density presented in Section \ref{chinferscheme}, 
 the parameters $\eta_2$, ..., $\eta_K$ are transformed to $z_2$, ..., $z_K$, where $\eta_k=\eta_1+\sum_{m=2}^{k}\sigma_{\eta}z_{m}$, $k=2,...,K$.
The conditional prior density of the $z$ parameters is $\pi(z_2, ...,z_K|\eta_{1},\sigma_{\eta}) = \prod_{k=2}^{K}\textrm{N}(z_k|0,1)=\prod_{k=2}^{K}\pi(z_k)$, i.e.,
that of independent Gaussian variates with mean zero and variance one.
  
Let $\psi_{5}=\eta_1$, and $\psi_{k+4}=z_k$, $k=2,...,K$. 
Then the prior density of $\fat{\psi}=(\psi_1,...,\psi_{K+4})$ is
$$
\pi(\fat{\psi}) = \prod_{k=1}^{K+4} \pi(\psi_k).
$$ 
The parameters in $\fat{\psi}$ are the hyperparameters of Model 1 and its latent parameters are $(\log(a),b,\fat{\beta})$, but recall that $b$ is set equal to $1.835$. 
The latent parameters of Model 0 are $(\log(a),b)$ and its hyperparameters are $(\psi_1,\psi_4,..., \psi_{K+4})$.

\subsection{Posterior sampling scheme}\label{chinferscheme} 

We propose a Markov chain Monte Carlo (MCMC) sampling scheme that is based on proposing the hyperparameters and the latent parameters jointly
to sample from the posterior density.
Our sampling scheme is motivated by the work of \citet{knorr2002block}. The posterior densities of Model 0 and Model 1 can be presented as
$$
\pi(\fat{x},\fat{\psi}|\fat{y}) \propto \pi(\fat{y}|\fat{x},\fat{\psi}) \pi(\fat{x}|\fat{\psi}) \pi(\fat{\psi})  
$$
where $\fat{y}=(\textcolor{black}{\tilde{Q}_1},...,\textcolor{black}{\tilde{Q}_n})\trp$ contains the discharge observations, and $\fat{x}$ and $\fat{\psi}$ denote the latent parameters and the hyperparameters, respectively,
of either Model 0 or Model 1.
The data density $\pi(\fat{y}|\fat{x},\fat{\psi})$ \textcolor{black}{is the product of lognormal densities} with location and scale parameters described by (\ref{eq:datadens})
and (\ref{eq:sigmah}), respectively, while $\pi(\fat{x}|\fat{\psi})$ denotes the Gaussian prior density of the latent parameters
under either Model 0 or Model 1 conditional on the hyperparameters.    

Under Model 1 let $A=(\fat{1} \ \fat{g} \ G)$ where $\fat{1}$ is a vector of ones,
$\fat{g}$ and $G$ are a vector and a diagonal matrix, respectively,
such that $\fat{g}_i=G_{ii}=\log(h_i-c)$, $i=1,...,n$. The prior
mean and covariance of $\fat{x}$ are $\fat{\mu}_x=(3.0, 1.835,\fat{0}\trp)\trp$ and  
$\Sigma_{x}=\textrm{bdiag}(3.0^2,0,\Sigma_{\beta})$, respectively, where
$\textrm{bdiag}$ denotes a block diagonal matrix and $\fat{0}$ is a vector of zeros. 
Under Model 0 then $A=(\fat{1} \ \fat{g})$, and the prior mean and covariance of $\fat{x}$  are $\fat{\mu}_x=(3.0, 1.835)\trp$ and  
$\Sigma_{x}=\textrm{diag}(3.0^2,0.426^2)$, respectively.
The diagonal matrix
$\Sigma_{\epsilon}$ is such that its $i$-th diagonal element is
$\sigma^2_{\epsilon}(h_i)$ and it is the same under Models 0 and 1.
The location and scale parameters of $\pi(\fat{y}|\fat{x},\fat{\psi})$
are $A\fat{x}$ and $\Sigma_{\epsilon}$, respectively.
 
The proposed MCMC sampling scheme is as follows.
The $k$-th posterior sample of $(\fat{x},\fat{\psi})$ is obtained by 
\begin{enumerate}[]
\item 
sampling $\fat{\psi}^{(k)}$  from $\pi(\fat{\psi}|\fat{y}) \propto \pi(\fat{\psi})\pi(\fat{y}|\fat{\psi})$
\item 
sampling $\fat{x}^{(k)}$  from $\pi(\fat{x}|\fat{y},\fat{\psi}^{(k)})$
\end{enumerate}
\noindent
and $\fat{\psi}^{(k)}$ and $\fat{x}^{(k)}$ are accepted jointly, so, if $\fat{\psi}^{(k)}$ is rejected, sampling of $\fat{x}^{(k)}$ can be delayed. \textcolor{black}{This is due to the fact that the acceptance ratio for a proposal $(\fat{\psi}^{*},\fat{x}^{*})$ is independent of $\fat{x}^{(k-1)}$ and $\fat{x}^{*}$ \citep[see][]{knorr2002block,GeHrSiSi2020}}.
The conditional posterior density $\pi(\fat{x}|\fat{y},\fat{\psi}^{(k)})$ is a multivariate Gaussian density derived from $\pi(\fat{y}|\fat{x},\fat{\psi}) \pi(\fat{x}|\fat{\psi})$. 
It has mean
$$
\fat{\mu}_{x|y} = \fat{\mu}_{x} - \Sigma_x A\trp (A\Sigma_x A\trp+\Sigma_\epsilon)^{-1} (A \fat{\mu}_x - \fat{v})
$$
where $\fat{v}=(\log(\textcolor{black}{\tilde{Q}_1}),...,\log(\textcolor{black}{\tilde{Q}_n}))$ and covariance
$$  
\Sigma_{x|y} = \Sigma_{x} - \Sigma_x A\trp (A\Sigma_x A\trp+\Sigma_\epsilon)^{-1}A \Sigma_x. 
$$
This is essentially the one block sampler of \citet{knorr2002block} in the case where the data density is lognormal.
The marginal posterior density $\pi(\fat{\psi}|\fat{y})$ is known up to a normalizing constant. The marginal density of $\fat{y}$, $\pi(\fat{y}|\fat{\psi})$, is lognormal with location and scale parameters
$A\fat{\mu}_x$ and $A\fat{\Sigma}_x A\trp+\Sigma_{\epsilon}$, respectively.
It is found by deriving the marginal distribution of $\fat{y}$ (conditional on $\fat{\psi}$) from the joint density of $(\fat{y},\fat{x})$, i.e., $\pi(\fat{y}|\fat{x},\fat{\psi}) \pi(\fat{x}|\fat{\psi})$. 

To obtain draws of $\fat{\psi}$, a random-walk Metropolis algorithm was used with a proposal density, $q(\cdot|\cdot)$, that is a multivariate Gaussian density centered on the last draw and with precision matrix $u^{-1}(-H)$, where $H$ is a finite difference estimate of the Hessian matrix of $\log(\pi(\fat{\psi}|\fat{y}))$ evaluated at the mode of $\log(\pi(\fat{\psi}|\fat{y}))$ with the mode denoted by $\hat{\fat{\psi}}$, also referred to as the posterior mode of $\fat{\psi}$.  
$H$ is given by
$$H \simeq \left. \nabla^2 \log(\pi(\fat{\psi}|\fat{y})) \right\vert_{\fat{\psi}=\hat{\fat{\psi}}}$$
and $u$ is a scaling constant given by $u=2.38^2/\text{dim}(\fat{\psi})$,
see \citet{Roberts1997}. 

\citet{Roberts1997} show that this scaling is optimal in a particular large dimension scenario. 
It turns out that this scaling works well for Model 0 and Model 1. 
Setting a specific scaling that is efficient removes the need for tuning.
The proposal value $\fat{\psi}$ in the $k$-th iteration given the draw of the $(k-1)$-th iteration, $\fat{\psi}^{(k-1)}$, is thus drawn from the following Gaussian density,
$$
\pi(\fat{\psi}|\fat{\psi}^{(k-1)}) = \textrm{N}(\fat{\psi}|\fat{\psi}^{(k-1)}, u(-H)^{-1}).
$$ 

\subsection{Prediction of discharge at observed and unobserved stage}\label{predictions} 

To sample from the posterior predictive distribution of discharge \textcolor{black}{$\tilde{Q}_i$} under Model 1 with stage $h_i$ found in the paired dataset then we sample
first $\fat{x}^{(l)}$ and $\fat{\psi}^{(l)}$ from the posterior density and then we draw a sample from
\begin{equation}\nonumber
\textcolor{black}{\tilde{Q}_i} \sim \textrm{LN}(\log(a)^{(l)}+\{b+\beta(h_i)^{(l)}\}\log(h_i-c^{(l)}),\sigma_\epsilon^2(h_i)^{(l)}).
\end{equation}
To draw a sample from the posterior predictive distribution of discharge \textcolor{black}{$\tilde{Q}_{\textrm{un}}$} corresponding unobserved stage $h_{\textrm{un}}$ under Model 1,
we use 
\begin{equation}\nonumber
\textcolor{black}{\tilde{Q}}_{\textrm{un}} \sim \textrm{LN}(\log(a)^{(l)}+\{b+\beta(h_{\textrm{un}})^{(l)}\}\log(h_{\textrm{un}} -c^{(l)}),\sigma_\epsilon^2(h_{\textrm{un}})^{(l)}),
\end{equation}
and $\beta(h_{\textrm{un}})^{(l)}$ is drawn from the conditional Gaussian density
$$
\pi(\beta(h_{\textrm{un}})|\fat{\beta}^{(l)},\psi_2^{(l)},\psi_3^{(l)}) = \textrm{N}(\beta(h_{\textrm{un}})|\fat{\gamma} \trp\Sigma_{\beta}^{-1}\fat{\beta}^{(l)},\sigma_{\beta}^2 - \fat{\gamma}\trp \Sigma_{\beta}\fat{\gamma}),
$$
where $\fat{\gamma}=\textrm{cov}(\fat{\beta},\beta(h_{\textrm{un}}))$ is the covariance between $\beta(h_{\textrm{un}})$ and the elements of $\fat{\beta}$, found from the Mat{\'e}rn covariance function in (\ref{eq:matcov}),
and $\fat{\gamma}$, $\Sigma_{\beta}$ and $\sigma^2_{\beta}$ are evaluated with $\psi_2=\psi_2^{(l)}$ and $\psi_3=\psi_3^{(l)}$. 
Samples from the posterior predictive distribution of discharge \textcolor{black}{$\tilde{Q}_0$} for observed or unobserved stage $h_0$ under Model 0 can be drawn from
\begin{equation}\nonumber
\textcolor{black}{\tilde{Q}}_0  \sim \textrm{LN}(\log(a)^{(l)}+b^{(l)}\log(h_0 -c^{(l)}),\sigma_\epsilon^2(h_0)^{(l)}),
\end{equation}
where the values of the parameters are equal to the $l$-th draw from the posterior distribution of $(\fat{x},\fat{\psi})$ under Model 0. 

\section{Results}\label{rebbs}
Here results based on a detailed analysis of \textcolor{black}{the data from} the four rivers introduced in Section \ref{chDATA} are given. This analysis was based on applying the two statistical models introduced in Section \ref{chModels} to the data and comparing these two models.

\subsection{\textcolor{black}{Computation and Convergence Assessment}}
\textcolor{black}{
Four chains were simulated for Model 0 and Model 1 where each chain consisted of \textcolor{black}{18,000} iterations and \textcolor{black}{2,000} burn-in iterations.  
This proved sufficient for all datasets. For both models a thinning factor of $5$ was used, meaning every $5$-th sample is kept for statistical inference and the rest discarded.
Running the four chains in parallel with code written in Matlab on an Intel(R) Core(TM) i5-7300U CPU (2.7GHz clock speed, 4 cores) with 16GB RAM simulations took $27$ seconds for Model 0 and $65$ seconds for Model 1 for the largest dataset ($86$ observations).}

\textcolor{black}{
Convergence in simulations from the posterior was ensured by assessing the Gelman-Rubin statistic \citep{Gelman1992,Gelman2013}, an estimate of a potential scale reduction factor, and by visually assessing trace plots, histograms and the autocorrelation function for a given simulated parameter (see the Supplementary Material).
The proposed posterior sampling schemes worked well judging from the Gelman--Rubin statistics, the autocorrelation function plots and visual inspection of the trace plots. The autocorrelation between samples $50$ iterations apart (no thinning applied) was at most $0.2$ and the Gelman--Rubin statistic was safely under reference bounds when the length of the chains after burn-in was greater than 10000.}

\subsection{Estimates of parameters and functions}
The two statistical models proposed in Section \ref{chModels} were applied to the datasets from the four rivers. Table \ref{Parm1} presents posterior summary statistics of selected parameter of these two
models for the four rivers. 
The estimates of the parameters $a$ and $c$ are somewhat different between the two models and the uncertainty in $a$ and $c$ in Model 1 is larger than in Model 0. The parameter $b$ in Model 0 is presented along with $b+\beta(2)$ in Model 1 (the power-law exponent at $h=2$ m) to provide comparison of the exponents of the two models. In the case of the Jokulsa a Fjollum River the $95$\% posterior intervals of these two quantities overlap the most and their widths are similar while for the other three rivers the $95$\% posterior intervals for $b+\beta(2)$ in Model 1 are wider by a factor \textcolor{black}{$1.7$} to \textcolor{black}{$3.1$} than those for for $b$ in Model 0.    

\begin{center}
\begin{table*}[hbt!] 
\caption{The 2.5\%, 50\% and 97.5\% posterior percentiles of the parameters $a$ (m$^3/$s), $b$ and $c$ (m) in Model 0 and the parameters $a$ (m$^3/$s), $b+\beta(2)$ and $c$ (m) in Model 1.}\label{Parm1}
\centering
\begin{tabular}{lcccccccc}
\toprule
\multicolumn{1}{c}{\textbf{ }} & \multicolumn{1}{c}{\textbf{ }} & \multicolumn{3}{c}{\textbf{Model 0}} & \multicolumn{1}{c}{\textbf{ }} & \multicolumn{3}{c}{\textbf{Model 1}} \\
\cmidrule(l{3pt}r{3pt}){3-5} \cmidrule(l{3pt}r{3pt}){7-9}
\textbf{River} & \textbf{Param.} & \textbf{2.5\%} & \textbf{50\%} & \textbf{97.5\%} & \textbf{Param.} & \textbf{2.5\%} & \textbf{50\%} & \textbf{97.5\%}\\
\midrule
Jokulsa a Dal & $a$ & 84.68 & 101.85 & 118.73 & $a$ & 34.61 & 73.87 & 99.96\\
 & $b$ & 1.71 & 1.85 & 2.01 & $b+\beta(2)$ & 1.91 & 2.20 & 2.85\\
 & $c$ & 0.61 & 0.70 & 0.78 & $c$ & 0.28 & 0.57 & 0.72\\
\midrule
Jokulsa a Fjollum & $a$ & 44.58 & 73.98 & 107.70 & $a$ & 52.91 & 94.38 & 119.52\\
 & $b$ & 1.84 & 2.09 & 2.40 & $b+\beta(2)$ & 1.75 & 1.93 & 2.31\\
 & $c$ & 0.06 & 0.32 & 0.52 & $c$ & 0.15 & 0.44 & 0.58\\
\midrule
Nordura & $a$ & 12.59 & 15.82 & 19.11 & $a$ & 14.30 & 19.42 & 22.82\\
 & $b$ & 1.97 & 2.15 & 2.31 & $b+\beta(2)$ & 1.63 & 1.85 & 2.22\\
 & $c$ & 0.80 & 0.89 & 0.97 & $c$ & 0.83 & 0.98 & 1.07\\
\midrule
Skjalfandafljot & $a$ & 3.80 & 6.38 & 10.03 & $a$ & 6.21 & 25.25 & 47.89\\
 & $b$ & 2.85 & 3.11 & 3.39 & $b+\beta(2)$ & 1.61 & 2.06 & 2.96\\
 & $c$ & -0.20 & 0.00 & 0.17 & $c$ & -0.08 & 0.55 & 0.91\\
\bottomrule
\end{tabular}
\end{table*}
\end{center}

\begin{figure}[hbt!]\centering
\begin{minipage}{0.9\linewidth}
\includegraphics[width=\linewidth]{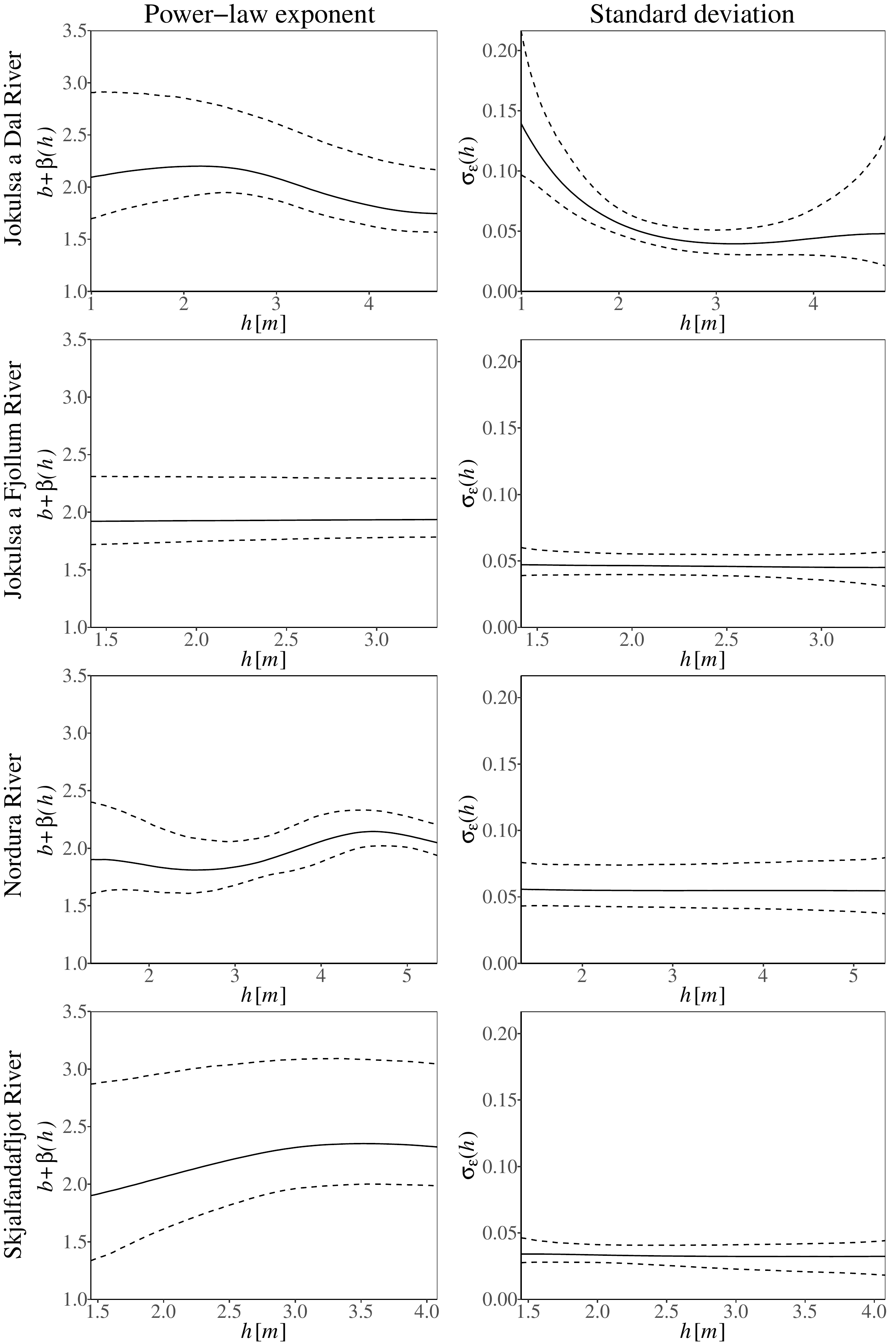}
\end{minipage}
\caption{Posterior estimate (---) and 95\% credible intervals (- -) for the power-law exponent $b+\beta(h)$ of Model 1 (left panel) and the standard deviation $\sigma_\epsilon(h)$ 
of Model 1 as function of stage, $h$, in meters (right panel).}\label{procsss}
\end{figure}

Figure \ref{procsss} displays the power-law exponent $b+\beta(h)$ and the standard deviation $\sigma_\epsilon(h)$ of Model 1 as a function of stage. \textcolor{black}{The power-law exponent, $b+\beta(h)$, of the Jokulsa a Fjollum River is close to being constant with respect to stage according to the posterior estimate, while it shows some variation with stage in the case of the other rivers.} The power-law exponent \textcolor{black}{can reveal} the geometry of the cross section of the corresponding river at the observational site. For example, in the case of the Jokulsa a Dal River, the posterior estimate of the process $b+\beta(h)$ takes values \textcolor{black}{slightly above} $2.0$ for low stage and for greater values of stage it takes values around $1.75$, \textcolor{black}{and is similar to $b+\beta(h)$ of the simulated river in the middle panel of Figure \ref{fig:irreg_cross_sec_fig}, indicating a parabolic-like shape for low stage and vertical river walls for higher stage}. The wide posterior intervals for $b+\beta(h)$ in the case of the Nordura River for values of stage \textcolor{black}{between $h=3.5$ m and $h=4.5$ m stem from the fact that none of the paired observations take stage values in this interval but they take stage values below and above the interval}.  
The posterior estimates of the standard deviation process $\sigma_\epsilon(h)$ 
indicate that it varies with stage in the case of the Jokulsa a Dal River while it is effectively constant for the other three rivers.

\begin{figure}[hbt!]\centering
\begin{minipage}{0.85\linewidth}
\includegraphics[width=\linewidth]{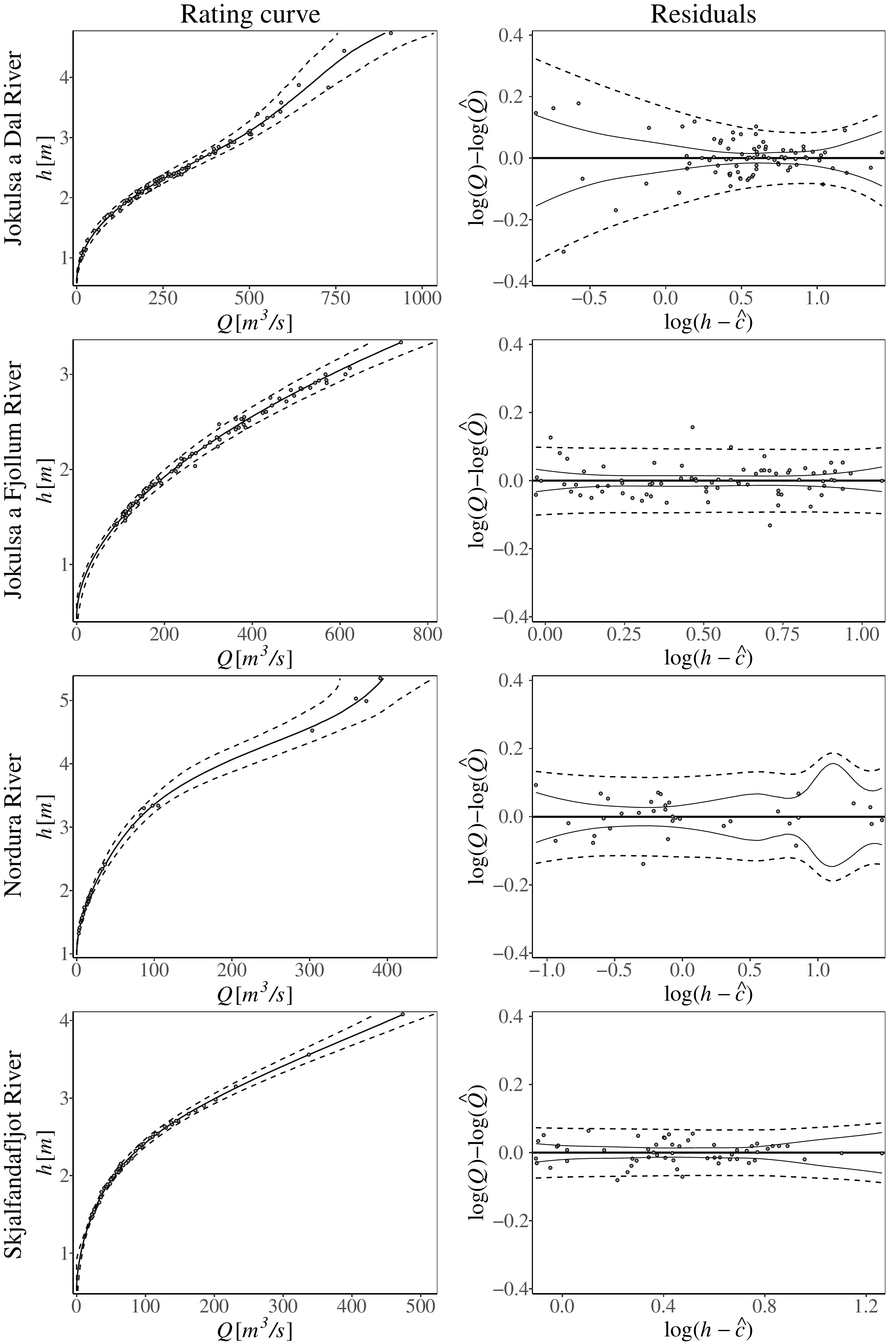}
\end{minipage}
\caption{Rating curves based on Model 1 for each of the four rivers are shown in the left panel. The vertical axis shows stage, $h$ (m), and the horizontal axis shows discharge, $Q$ (m$^3$/s). Estimates of rating curves (---) and 95\% predictive intervals (- -) are shown.
The residual plots for each of the four rivers are shown in the right panel. The vertical axis shows the log-residuals, i.e., the observed log-discharge minus the posterior estimate of log-discharge, denoted by $\log(\hat{Q})$. The horizontal axis shows the logarithm of estimated corrected stage. The residuals plots also show 95\% predictive intervals for $\log(Q)$ (- -) and 95\% credible intervals for the expected value of $\log(Q)$ (---), reflecting the rating curve uncertainty. The posterior estimate $\log(\hat{Q})$ is subtracted from these two types of intervals so that they can be plotted along with the residuals.}
\label{RCnormpepi}
\end{figure}

The left panel of Figure \ref{RCnormpepi} shows estimates of rating curves and predictive intervals for the four rivers under Model 1. Note that stage is shown on the vertical axes and discharge on the horizontal axes as this is the standard practice in hydrology. The generalized rating curves provided a convincing fit to the four datasets.  
Residual plots are presented in the right panel of 
Figure \ref{RCnormpepi} along with the prediction intervals and credible intervals for 
expected value of $\log(Q)$ on the same scale, but with the prediction estimates subtracted for better visualization.
The residual plots indicate that the mean of Model 1 captures the underlying mean and that the standard deviation of Model 1 describes the variability in the measurement errors as a function of stage adequately well.
The wide posterior predictive intervals \textcolor{black}{in the case of the Nordura River in the range from $h=3.5$ m to $4.5$ m are due to the absence of observations in this range of stage values}. 

\subsection{Model comparison}

Three statistics were used to compare Model 0 and Model 1. These three statistics are the deviance information criterion (DIC), posterior model probabilities (based on Bayes factor) and the average absolute prediction error in a leave-one-out cross-validation. These statistics are described below.

DIC quantifies the fit of a model to data with respect to the complexity of the model \citep{spiegelhalter2002bayesian}. It is given by
$$
{\rm DIC} = D_\text{avg} + p_D
$$
where $D_\text{avg}$ measures the fit of the model to the data. It is estimated with  
$$
\hat{D}_\text{avg} = -\frac{2}{L}\sum_{l=1}^{L}\log \pi(\boldsymbol{y}|\boldsymbol{\xi}^{(l)})
$$
where $\pi(\boldsymbol{y}|\boldsymbol{\xi}^{(l)})$ is the likelihood function which arises from the 
proposed probability model of the data and $\boldsymbol{\xi}^{(l)}=(\fat{x}^{(l)},\fat{\psi}^{(l)})$ is the $l$-th posterior sample of the parameters of the model. Lower values of $D_\text{avg}$ indicate \textcolor{black}{a better fit to the data}. $p_D$ measures the complexity of the model and is referred to as the effective number of parameters. It is usually less then the actual number of parameters, but these two numbers can be equal in some cases. 
$p_D$ is computed with $p_D = D_\text{avg} - D_{\hat\psi}$ where $D_{\hat\psi}$ is another measure of fit that is given by $-2\log \pi(\boldsymbol{y}|\hat{\boldsymbol{\xi}})$\textcolor{black}{, here $\hat{\boldsymbol{\xi}}$ contains the posterior median of each parameter}. DIC presents a trade-off between \textcolor{black}{the} fit of \textcolor{black}{a} model to the data and the model\textcolor{black}{'s} complexity, and it allows for comparison between models of different complexity in terms of the same dataset. 

The posterior probabilities of Model 0 and Model 1 were computed using the Bayes factor and assuming a priori that the two models were equally likely \citep[see][]{jeffreys1961theory,kass1995bayes}. 
The posterior probability of Model 1 when comparing it to Model 0 is given by
$$
{\rm P}(M_1|\boldsymbol{y}) = \frac{{\rm P}(M_1)\int_{H_1}\pi_1(\boldsymbol{y}|\boldsymbol{\xi}_1)\pi(\boldsymbol{\xi}_1) d \boldsymbol{\xi}_1}
{\sum_{s=0,1}{\rm P}(M_s)\int_{H_s}\pi_s(\boldsymbol{y}|\boldsymbol{\xi}_s)\pi(\boldsymbol{\xi}_s) d \boldsymbol{\xi}_s}
$$  
$$
=\left(1+\frac{{\rm P}(M_0)}{{\rm P}(M_1)}\times \frac{1}{B_{10}}    \right)^{-1}
$$
where $B_{10}$ is the Bayes factor for the comparison of models $M_1$
and $M_0$ \citep[see][]{kass1995bayes}. $B_{10}$ is given by
$$
B_{10}= \frac{\int_{H_1}\pi_1(\boldsymbol{y}|\boldsymbol{\xi}_1)\pi(\boldsymbol{\xi}_1) d \boldsymbol{\xi}_1}{\int_{H_0}\pi_0(\boldsymbol{y}|\boldsymbol{\xi}_0)\pi(\boldsymbol{\xi}_0) d \boldsymbol{\xi}_0}
$$
and \textcolor{black}{here} it is computed by approximating the integrals $\int_{H_s}\pi_s(\boldsymbol{y}|\boldsymbol{\xi}_s)\pi(\boldsymbol{\xi}_s) d \boldsymbol{\xi}_s$, $s=0,1$, with
\textcolor{black}{the harmonic mean of the likelihood values, i.e.,}
$$
\left\{ \frac{1}{L} \sum_{l=1}^{L} \frac{1}{\pi_s(\boldsymbol{y}|\boldsymbol{\xi}_s^{(l)})}  \right\}^{-1}
$$
\textcolor{black}{where $\boldsymbol{\xi}_s^{(l)}$ is the $l$-th posterior sample of $\boldsymbol{\xi}_s$ in Model $s$, see \cite{newton1994approximate} for details. However, in some cases the theoretical variance of the harmonic mean estimator is not finite \citep{newton1994approximate}, and even when it is finite, it can become very large since a small fraction of the posterior samples can correspond to a small likelihood which will have a large effect on the harmonic mean estimate. Furthermore, the harmonic mean estimator can be biased even though it is asymptotically unbiased \citep{raftery2007estimating,CALDERHEAD20094028}. Thus, here, the posterior model probabilities are treated with caution and viewed in light of the other model comparison statistics. Additionally, a simulation study is provided in Appendix \ref{appC} to assess the variability in the estimated posterior model probabilities and in the difference in DIC values.}

A leave-one-out cross-validation was applied to both Model 0 and Model 1. For each dataset and Model $s$, one of the observations was left out and then the expected value of $\log(Q)$ was predicted for that stage using an Empirical Bayes type approximation to save computation, i.e., the posterior mode of the hyperparameters was used but their uncertainty was not taken into account. The absolute value of the difference between the logarithm of the observed discharge and the prediction was computed. This was done for all of the observations within the dataset. Finally, the average of these absolute differences, $\overline{\rm AD}_{\rm{cv},i}$, was computed.

\begin{center}
\begin{table*}[ht!]
\caption{Comparison of Model 0 and Model 1. Here $P_s=\textrm{P}$(Model $s$|Data), $s=0,1$, and $\overline{\textrm{AD}}_{\textrm{cv},s}$ is the average absolute prediction error between a prediction based on Model $s$ and the left out observation (log-scale) in a leave-one-out cross-validation. 
A description of the quantities $D_{\hat\theta}$, $\hat{D}_\textrm{avg}$, DIC, $\Delta$DIC, $p_d$ and $\overline{\textrm{AD}}_{\textrm{ratio}}$ is given in the main text.}\label{MComp}
\centering
\begin{tabular}{lccccccccc}
\toprule
\textbf{River} & \textbf{Model} & $D_{\hat{\theta}}$ & $\hat{D}_{\textrm{avg}}$ & DIC & $\Delta$DIC & $p_d$ & $P_s$ & $\overline{\textnormal{AD}}_{\textrm{cv},s}$ & $\overline{\textnormal{AD}}_{\textrm{ratio}}$\\
\midrule
Jokulsa a Dal & 0 & 714.1 & 719.7 & 725.4 & 49.9 & 5.7 & 0.0000 & 0.0785 & 1.501\\
 & 1 & 659.0 & 667.2 & 675.5 &  & 8.3 & 1.0000 & 0.0523 & \\
\midrule
Jokulsa a Fjollum & 0 & 586.6 & 589.8 & 592.9 & -0.8 & 3.2 & 0.5028 & 0.0342 & 0.991\\
 & 1 & 588.0 & 590.9 & 593.7 &  & 2.8 & 0.4972 & 0.0345 & \\
\midrule
Nordura & 0 & 135.4 & 138.5 & 141.6 & 23.2 & 3.1 & 0.0000 & 0.0725 & 1.102\\
 & 1 & 104.8 & 111.6 & 118.4 &  & 6.8 & 1.0000 & 0.0658 & \\
\midrule
Skjalfandafljot & 0 & 264.3 & 267.3 & 270.3 & 14.2 & 3.0 & 0.0007 & 0.0330 & 1.196\\
 & 1 & 246.1 & 251.1 & 256.0 &  & 5.0 & 0.9993 & 0.0276 & \\
\bottomrule
\end{tabular}
\end{table*}
\end{center}


Table \ref{MComp} shows the above model comparison statistics; $D_{\hat\theta}$; $\hat{D}_\text{avg}$; DIC; the DIC difference between Model 0 and Model 1 given by $\Delta$DIC$=$DIC$_0-$DIC$_1$; $p_d$; posterior probabilities of the two models when assuming equal prior probabilities; $\overline{\rm AD}_{\rm{cv},s}$; the ratio $\overline{\rm AD}_{\rm{ratio}}=\overline{\rm AD}_{\rm{cv},0}/\overline{\rm AD}_{\rm{cv},1}$. 
In the case of the Jokulsa a Fjollum River, Model 0 seems to fit well, which is consistent with the fact that the power-law exponent in Model 1 was close to being constant with respect to stage. Model 1 gives a slightly better fit according to $D_{\hat\theta}$ while Model 0 has a smaller DIC and a smaller effective number of parameters. The posterior probability of Model 1 when compared to Model 0  \textcolor{black}{is close to $0.5$ (uncertainty bounds $(0.12;0.73$), see Appendix \ref{appC})}, indicating that Model 0 and Model 1 fit the data equally well. Also, along these lines, $\overline{\rm AD}_{\rm{ratio}}$ is very close to one or $0.991$, indicating equal performance of the two models. Considering the small differences between the two models in terms of DIC, the posterior model probabilities and $\overline{\rm AD}_{\rm{cv}}$, it seems reasonable to assume them to be equally good.  
This implies that when Model 0 is adequate then Model 1 can mimic Model 0. The Bayesian hierarchical model behind Model 1 is designed to have Model 0 as special case and this example proves that this feature of Model 1 works in practice.

In the case of the Skjalfandafljot River, the variation in the $b+\beta(h)$ process in Model 1 was modest but it yielded a better fit than Model 0 resulting in a large DIC difference, $\overline{\rm AD}_{\rm{ratio}}$ of 1.196 and posterior model probability decisively favoring Model 1. When Model 1 was applied to the Nordura River, the variation in the $b+\beta(h)$ process was high and Model 1 yielded a better fit\textcolor{black}{, the} DIC difference and $\overline{\rm AD}_{\rm{ratio}}$ were large and Model 1 was decisively favored according to the posterior model probability. For the Jokulsa a Dal River, the misfit of Model 0 was most obvious and Model 1 had a decent variation in the $b+\beta(h)$ process, the DIC difference was large,  $\overline{\rm AD}_{\rm{ratio}}$ was large and posterior model probability favored Model 1 decisively.

\section{Conclusion}\label{chConclusion}
By linking the physics formulas of Ch{\'e}zy and Manning for open channel flow to a Bayesian hierarchical model, we obtain a flexible class of rating curves. We explored the properties of these curves, which are referred to as generalized power-law rating curves. 
The power-law exponent of the generalized power-law rating curve, $f(h)$, 
was explored through cross sections of open channels with relatively simple geometry as well as more irregular forms closer to what one might expect to find in nature.
This exploration revealed that $f(h)$ takes values in a relatively narrow range, namely, the interval $[1.0,2.67]$.
This is an important fact when constructing prior densities for the unknown parameters of the Bayesian hierarchical model that are linked to $f(h)$.

The model for the power-law exponent assumes that $f(h)$ is continuous and $f'(h)$ is piecewise continuous. In terms of statistical inference, $f(h)$ is not observed directly and subject to measurement error. There is little information to detect small scale rapid changes in $f(h)$ while smoother changes in $f(h)$ over a wider range in $h$ can be detected. 
Therefore we assumed that $f(h)$, $f'(h)$ and $f''(h)$ are continuous within the Bayesian hierarchical model. To fulfil these assumptions and leave room for a certain amount of flexibility, $f(h)$ is modeled as a two times mean square differentiable Gaussian process of a Mat{\'e}rn type.

We proposed an efficient MCMC sampling scheme in line with that of \citet{knorr2002block} for the Bayesian hierarchical model. It is efficient in terms of the autocorrelation decaying reasonably fast and the Gelman--Rubin statistics researching its convergence reference value after a relatively small number of sampling iterations. The sampling scheme utilizes the structure of the Bayesian hierarchical model, namely, that lognormal and Gaussian distributions are assumed at the data level and the latent level, respectively. The hyperparameters are sampled from their marginal posterior density while the latent parameters are sampled from their conditional density, which is Gaussian. Due to this set-up, the efficiency of the sampling scheme depends on the efficiency of the sampling scheme for the marginal posterior density of the hyperparameters. The sampling approach of \citet{Roberts1997} was selected for the hyperparameters. This sampling approach worked well here. Transforming the random walk variates in the variance model to standardized variates improved the geometry of the marginal posterior density of the hyperparameters and led to a stable sampling scheme.  

Generalized power-law rating curves and power-law rating curves were applied to four real stage-discharge datasets, and their performance was evaluated. Our analysis showed that data of this type can be modeled appropriately with the generalized power-law rating curve, and that in some cases the power-law rating curve gives a sufficient description of the data. This demonstrates that setting forth a statistical model that is motivated by the hydrodynamic theory summarized by the formulas of Manning and Ch{\'e}zy is a logical approach for the estimation of rating curves. Furthermore, the model comparison based DIC, posterior model probability (Bayes factor) and the average absolute prediction error in a leave-one-out cross-validation, can be used to determine whether Model 1 is more appropriate than Model 0 or whether Model 0 is sufficient.
In the case where the power-law rating curve performed better than the generalized power-law rating curve, the fit of the generalized power-law rating curve was convincing, mimicking the power-law rating curve though with wider posterior intervals. This is to be expected as, in theory, the generalized power-law rating curve is flexible enough to reduce to the power-law rating curve when the data point in that direction. 

The Bayesian hierarchical model for the generalized power-law rating curve provides a novel approach for fitting rating curves. In the analysis conducted here it has been proven to be flexible enough to provide good fit to the data, and it is supported by an efficient MCMC sampling scheme. 


\section*{Acknowledgments}
The authors would like to express thanks to the Landsvirkjun Energy Research Fund and the Research Fund of Vegagerdin that supported this research. 
The authors also express their thanks to the Nordic Network on Statistical Approaches to Regional Climate Models for Adaptation (SARMA) for providing travel support. 




%
%

\appendix

\section{Hydrodynamic quantities and parameters\label{app0}}

Table \ref{hydroquanties} contains a list of the hydrodynamic quantities and parameters found in Section \ref{sec:meanvelo} along with their units.

\begin{center}
\begin{table*}[ht!] 
\caption{A list of the hydrodynamic quantities and parameters found in Section \ref{sec:meanvelo} along with their units.}\label{hydroquanties}
\centering
\begin{tabular*}{336pt}{c l c}
\toprule
\textbf{Quantity} & \textbf{Description} & \textbf{Units}  \\ 
\midrule
$A$ & cross section area & m$^2$ \\
$P$ & wetted perimeter & m \\
$R$ & hydraulic radius & m \\
$S$ & slope of a channel & unit free \\
$\bar{v}$ & mean velocity through a cross section & m$/$s \\ 
$Q$ & discharge  &  m$^3/$s  \\
$C$ & Chezy's constant & m$^{1/2}$ s$^{-1}$ \\
$n$ & Manning's constant & s m$^{-1/3}$  \\
$f$ & the Darcy--Weisbach friction factor & unit free  \\
$g$ & the Earth's gravitational acceleration & m$/$s$^{-2}$ \\
$h$ & stage (water elevation) or depth & m \\ 
$w(h)$ & the width of the cross-section at depth $h$ & m \\
$x$ & the exponent of $R$ in the mean velocity formula & unit free \\
$k$, $K_1$, $K_2$ & constants independent of stage &   \\   
\bottomrule
\end{tabular*}
\end{table*}
\end{center}

\section{Proofs of Results}

\subsection{Proof of Result 1\label{app1}}
\begin{proof}[Proof of Result 1]
Given the constants $k$ and $x$, and the continuous functions $A(h)$ and $P(h)$, 
assume there is a constant $a$ and a function $f(h)$ such that
$$
Q(h) = a h^{f(h)} = k A(h)^{x+1}P(h)^{-x}.
$$
In the case of $h=1$ then 
$$
Q(1) = a 1^{f(1)} = a = k A(1)^{x+1}P(1)^{-x},
$$
giving the result for $a$. This yields
$$
\log(a) - \log(k) = (x+1)\log\{A(1)\}-x\log\{P(1)\}.
$$
Take the logarithm of the two forms for $Q(h)$ to obtain
$$
\log(a) + f(h)\log(h) = \log(k) + (x+1)\log\{A(h)\}-x\log\{P(h)\}
$$
and
$$
f(h)\log(h) = -\log(a) + \log(k) + (x+1)\log\{A(h)\}-x\log\{P(h)\}
$$
$$
=  - (x+1)\log\{A(1)\}+x\log\{P(1)\} + (x+1)\log\{A(h)\}-x\log\{P(h)\}.
$$
Assume that $\log(h) \neq 0$ then
$$
f(h)=\frac{ (x+1)\log
\left\{\displaystyle\frac{A(h)}{A(1)}\right\}
-x\log
\left\{\displaystyle\frac{P(h)}{P(1)}\right\}}{\log(h)}
$$
which gives (\ref{eq:gplc}).
\end{proof}

\subsection{Proof of Result 2\label{app2}}
\begin{proof}[Proof of Result 2]

Note that
$$
A'(h) = \frac{\textrm{d}}{\textrm{d}h} A(h) = \frac{\textrm{d}}{\textrm{d}h} \int_{0}^{h}(w_1(\eta)+w_2(\eta)) d\eta = w_1(\eta)+w_2(\eta),
$$
$$
A''(h) = \frac{\textrm{d}}{\textrm{d}h} A'(h) = w_1'(\eta)+w_2'(\eta),
$$
$$
P'(h) = \frac{\textrm{d}}{\textrm{d}h} P(h) = 
\frac{\textrm{d}}{\textrm{d}h} \int_{0}^{h} \sqrt{1 + \{w_1'(\eta)\}^2 }  d\eta +
\frac{\textrm{d}}{\textrm{d}h} \int_{0}^{h} \sqrt{1 + \{w_2'(\eta)\}^2 }  d\eta
$$
$$
=\sqrt{1 + \{w_1'(h)\}^2 }+\sqrt{1 + \{w_2'(h)\}^2 },
$$
$$
P''(h) = \frac{\textrm{d}}{\textrm{d}h} P'(h) = \frac{\textrm{d}}{\textrm{d}h} \sqrt{1 + \{w_1'(h)\}^2 }
+\frac{\textrm{d}}{\textrm{d}h} \sqrt{1 + \{w_2'(h)\}^2 }
$$
$$
= \frac{w_1'(h)w_1''(h)}{\sqrt{1 + \{w_1'(h)\}^2 }}
+ \frac{w_2'(h)w_2''(h)}{\sqrt{1 + \{w_2'(h)\}^2 }}.
$$

The limit of $f(h)$ as $h$ approaches $1$ can be evaluated using L'H\^opital's rule once. That is,
$$
\lim_{h \rightarrow 1} f(h) =
\lim_{h \rightarrow 1}
\frac{(x+1)\{\log A(h) - \log A(1)\} - x\{\log P(h) - x\log P(1)\}}{\log h}
$$
$$
=\lim_{h \rightarrow 1}
\frac{(x+1)A'(h) A(h)^{-1} - x P'(h) P(h)^{-1}}{h^{-1}}
$$
$$
=(x+1) \frac{A'(1)}{A(1)} - x \frac{P'(1)}{P(1)},
$$
which gives (\ref{eq:gplb3}).
The limits of $f(h)$ as $h$ approaches zero from above and infinity can be evaluated using
L'H\^opital's rule twice. In the case of the limit where $h$ approaches zero from above then after applying L'H\^opital's rule once then
$$
\lim_{h \rightarrow 0^{+}}f(h)=
\lim_{h \rightarrow 0^{+}}
\frac{(x+1)A'(h) A(h)^{-1} - x P'(h) P(h)^{-1}}{h^{-1}}
$$
$$
=(x+1)\lim_{h \rightarrow 0^{+}} \frac{h A'(h)}{A(h)}
- x\lim_{h \rightarrow 0^{+}} \frac{h P'(h)}{P(h)}
$$  
$$
=(x+1)\lim_{h \rightarrow 0^{+}} \frac{\{A'(h) + hA''(h) \}}{A'(h)}
- x\lim_{h \rightarrow 0^{+}} \frac{ \{P'(h) + hP''(h) \}}{P'(h)}
$$
$$
=(x+1)\lim_{h \rightarrow 0^{+}}\left\{1+ \frac{hA''(h)}{A'(h)} \right\}
- x\lim_{h \rightarrow 0^{+}} \left\{1+ \frac{hP''(h)}{P'(h)} \right\}
$$
$$
=1 +
(x+1)\lim_{h \rightarrow 0^{+}} \frac{hA''(h)}{A'(h)} 
- x\lim_{h \rightarrow 0^{+}}  \frac{hP''(h)}{P'(h)},
$$
which gives (\ref{eq:gplb2}).

The proof for the limit of $f(h)$ as $h$ approaches infinity, which is given in (\ref{eq:gplb4}), is similar as it is also based on applying L'H\^opital's rule twice. The main difference is that in the case of $h$ approaching infinity the functions in both the numerator and the dominator approach infinity while in the case of $h$ approaching zero from above the functions in both the numerator and the dominator approach zero; however, L'H\^opital's rule is applicable in both cases.  
\end{proof}

\subsection{Proof of Result 3\label{app3}}
\begin{proof}[Proof of Result 3]

Note that
$$
A_{\alpha}'(h) = w_{\alpha}(h) = \phi_{\alpha}h^{\alpha}, \quad
A_{\alpha}''(h) =  w_{\alpha}'(h) = \alpha\phi_{\alpha}h^{\alpha-1},
$$
$$
P_{\alpha}'(h)  
= 2 \sqrt{1 + 4^{-1}\{w_{\alpha}'(h)\}^2 }  
= 2 \sqrt{1 + 4^{-1}\alpha^2\phi^2_{\alpha} h^{2\alpha-2} },
$$
$$
P_{\alpha}''(h) 
= \frac{w_{\alpha}'(h)w_{\alpha}''(h)}
{2 \sqrt{1 + 4^{-1}\{w_{\alpha}'(h)\}^2 }}
=\frac{\alpha\phi_{\alpha}h^{\alpha-1} (\alpha-1)\alpha\phi_{\alpha}h^{\alpha-2}}
{2 \sqrt{1 + 4^{-1}\alpha^2\phi^2_{\alpha} h^{2\alpha-2} }}
=\frac{(\alpha-1) \alpha^2 \phi_{\alpha}^2 h^{2\alpha-3}}
{2 \sqrt{1 + 4^{-1}\alpha^2\phi^2_{\alpha} h^{2\alpha-2} }}.
$$
As
$$
\frac{A_{\alpha}(h)}{A_{\alpha}(1)} = h^{\alpha+1}
$$
then according to (\ref{eq:gplc}) the form of $f_{\alpha}(h)$ becomes
$$
f_{\alpha}(h) = 
\frac
{(x+1)\log(h^{\alpha+1}) - x\log\{P_{\alpha}(h)\} + x\log\{P_{\alpha}(1)\}}
{\log(h)}
$$
$$
= (x+1)(\alpha+1) 
-x\frac{[\log\{P_{\alpha}(h)\} - \log\{P_{\alpha}(1)\} ]}{\log(h)}.
$$
which gives (\ref{eq:gold}).
As 
$$
\frac{A_{\alpha}'(1)}{A_{\alpha}(1)} = (\alpha+1)
$$
then according to (\ref{eq:gplb3}) the limit of $f_{\alpha}(h)$ as $h$ approaches one is
$$
\lim_{h\rightarrow 1}f_{\alpha}(h)=
(x+1)\frac{A'_{\alpha}(1)}{A_{\alpha}(1)}-x\frac{P'_{\alpha}(1)}{P_{\alpha}(1)}
=(x+1)(\alpha+1)-x\frac{P'_{\alpha}(1)}{P_{\alpha}(1)},
$$
which gives (\ref{eq:gple}).
To evaluate the limit of $f_{\alpha}(h)$ as $h$ approaches zero from above note that
$$
\frac{hA_{\alpha}''(h)}{A_{\alpha}'(h)} = \frac{h \alpha \phi_{\alpha}h^{\alpha-1}}{\phi_{\alpha}h^{\alpha}} = \alpha 
$$
and that
$$
\frac{h P_{\alpha}''(h)}{P_{\alpha}'(h)} 
= h P_{\alpha}''(h)\times \frac{1}{P_{\alpha}'(h)}
=
\frac{h (\alpha-1) \alpha^2 \phi_{\alpha}^2 h^{2\alpha-3}}
{2 \sqrt{1 + 4^{-1}\alpha^2\phi^2_{\alpha} h^{2\alpha-2} }}
\times
\frac{1}{2 \sqrt{1 + 4^{-1}\alpha^2\phi^2_{\alpha} h^{2\alpha-2} }}
$$
$$
=(\alpha-1)\frac{4^{-1}\alpha^2  \phi_{\alpha}^2 h^{2\alpha-2}}
{1 + 4^{-1}\alpha^2\phi^2_{\alpha} h^{2\alpha-2} }.
$$
According to (\ref{eq:gplb2})
$$
\lim_{h\rightarrow 0^{+}} f_{\alpha}(h) = 1 + (x+1)\lim_{h\rightarrow 0^{+}}\frac{hA_{\alpha}''(h)}{A_{\alpha}'(h)} 
- x \lim_{h\rightarrow 0^{+}} \frac{h P_{\alpha}''(h)}{P_{\alpha}'(h)}
$$
$$
=1 + (x+1)\alpha - x \lim_{h\rightarrow 0^{+}}
(\alpha-1)\frac{4^{-1}\alpha^2  \phi_{\alpha}^2 h^{2\alpha-2}}
{1 + 4^{-1}\alpha^2\phi^2_{\alpha} h^{2\alpha-2} }
$$
which in the case of $0 \leq \alpha < 1$ becomes
$$
\lim_{h\rightarrow 0^{+}} f_{\alpha}(h) = \alpha + x + 1
$$
and in the case of $\alpha > 1$ this limit becomes
$$
\lim_{h\rightarrow 0^{+}} f_{\alpha}(h) = \alpha + 1 + \alpha x,
$$
which gives (\ref{eq:gpld}).
Same arguments can be used to show that the limit of $f_{\alpha}(h)$ as $h$ approaches infinity in the case of $0 \leq \alpha < 1$ becomes
$$
\lim_{h\rightarrow \infty} f_{\alpha}(h) = \alpha + 1 + \alpha x
$$
and in the case of $\alpha > 1$ the limit becomes
$$
\lim_{h\rightarrow \infty} f_{\alpha}(h) = \alpha + x + 1,
$$
which gives (\ref{eq:gplf}).
It was shown in the main text that the special case $\alpha = 1$ resulted in a constant exponent, that is, $f_{1}(h) = 2 + x$ for $h \geq 0$.
\end{proof}

\section{\textcolor{black}{The uncertainty in model comparison\label{appC}}}

\textcolor{black}{Here we present the uncertainty in two model comparison statistics, namely, the model probability of Model 1, $P_1$, and the DIC difference between Model 0 and Model 1, $\Delta$DIC, for each of the four datasets. These two statistics were computed $500$ times where each evaluation was based on a full MCMC run. Table \ref{MUncertainty} shows the $90$\% empirical intervals and medians of these two model comparison statistics based on the $500$ MCMC runs. The variability in the model probability is moderate when its value is close to $1.0$, however, when two models give a similar fit to the data then the variance of the harmonic mean estimator for the Bayes factor is apparent and has a substantial effect on the model probability. The variability in $\Delta$DIC is moderate.}

\begin{center}
\begin{table}[ht]
\caption{\textcolor{black}{The $90$\% empirical intervals and medians of (i) $P_1$, the model probability of Model 1; (ii) $\Delta$DIC, the DIC difference between Model 0 and Model 1; for each of the four datasets. The empirical intervals based on $500$ MCMC runs.}}\label{MUncertainty}
\centering
\begin{tabular}{lrrrrrr}
  \hline
\textbf{River} & $P_{1,\textrm{low}}$ & $P_{1,\textrm{med}}$ & $P_{1,\textrm{upp}}$ & $\Delta$DIC$_{\textrm{low}}$ & $\Delta$DIC$_{\textrm{med}}$ & $\Delta$DIC$_{\textrm{upp}}$ \\
  \hline
Jokulsa a Dal & $1.000000$ & $1.000000$ & $1.000000$ & $49.110664$ & $49.577520$ & $50.018969$ \\
  Jokulsa a Fjollum & $0.123694$ & $0.359365$ & $0.728100$ & $-1.004389$ & $-0.787565$ & $-0.555558$ \\
  Nordura & $0.998698$ & $0.999946$ & $0.999993$ & $22.981757$ & $23.298375$ & $23.583573$ \\
  Skjalfandafljot & $0.992445$ & $0.999619$ & $0.999957$ & $14.185779$ & $14.430806$ & $14.684749$ \\
   \hline
\end{tabular}
\end{table}
\end{center}

\bibliographystyle{aps-nameyear} 
\newcommand{\noopsort}[1]{} \newcommand{\printfirst}[2]{#1}
  \newcommand{\singleletter}[1]{#1} \newcommand{\switchargs}[2]{#2#1}

\end{document}


\baselineskip12pt

\maketitle
\section{Convergence assessment}\label{sm_sec1}

\begin{figure}[hbt!]\centering
\begin{minipage}{0.95\linewidth}
\includegraphics[width=\linewidth]{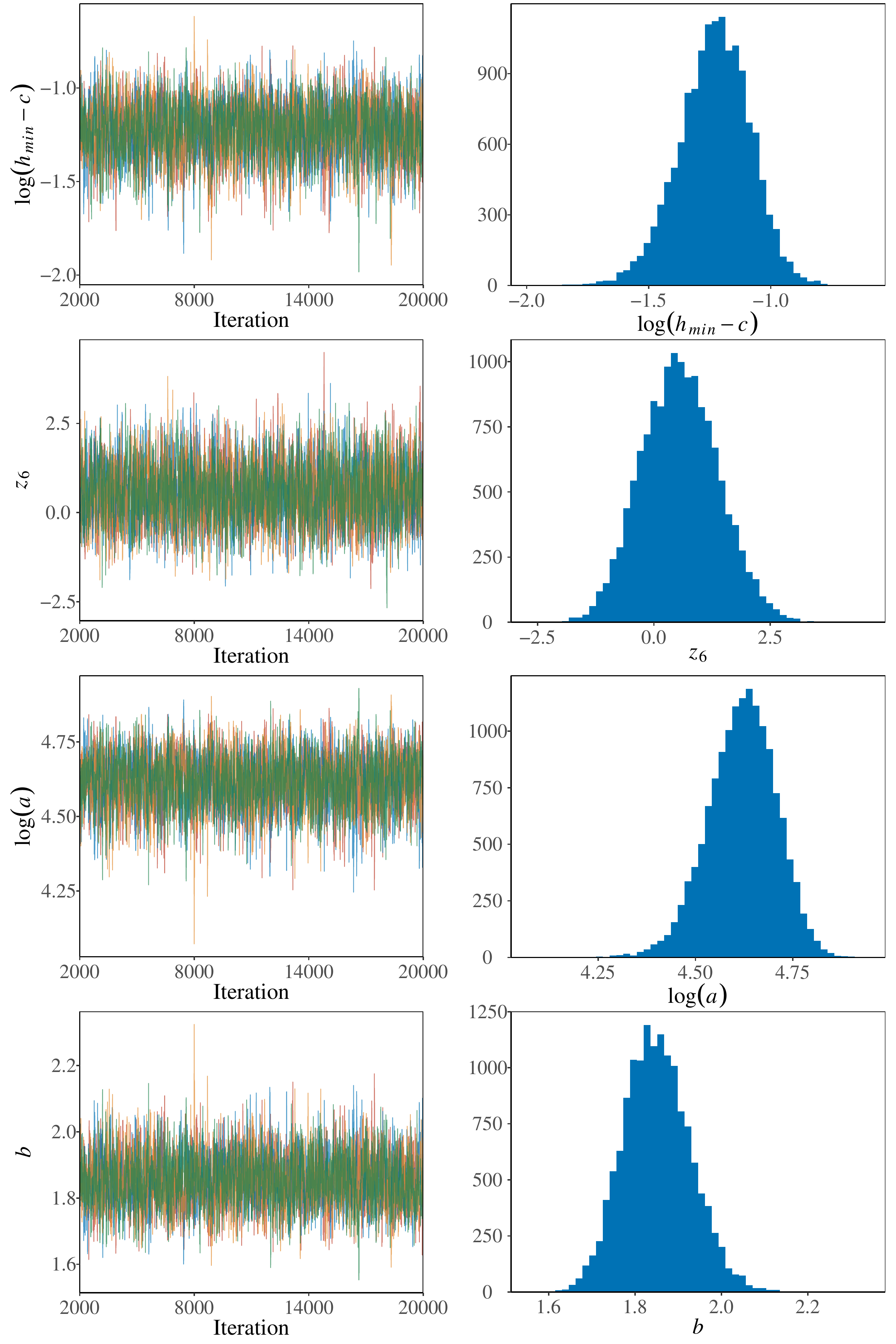}
\end{minipage}
\caption{Convergence diagnostics for Model 0 using the Jokulsa a Fjollum River dataset. Trace plots (left panel) of the four chains of simulation, and histograms (right panel) of samples from simulation for the following parameters;
$\psi_{1} = \log(h_{\min}-c)$; $\psi_{10} = z_6$; 
$x_1=\log(a)$; $x_2=b$.}\label{trhim1}
\end{figure}

\begin{figure}[hbt!]\centering
\begin{minipage}{0.95\linewidth}
\includegraphics[width=\linewidth]{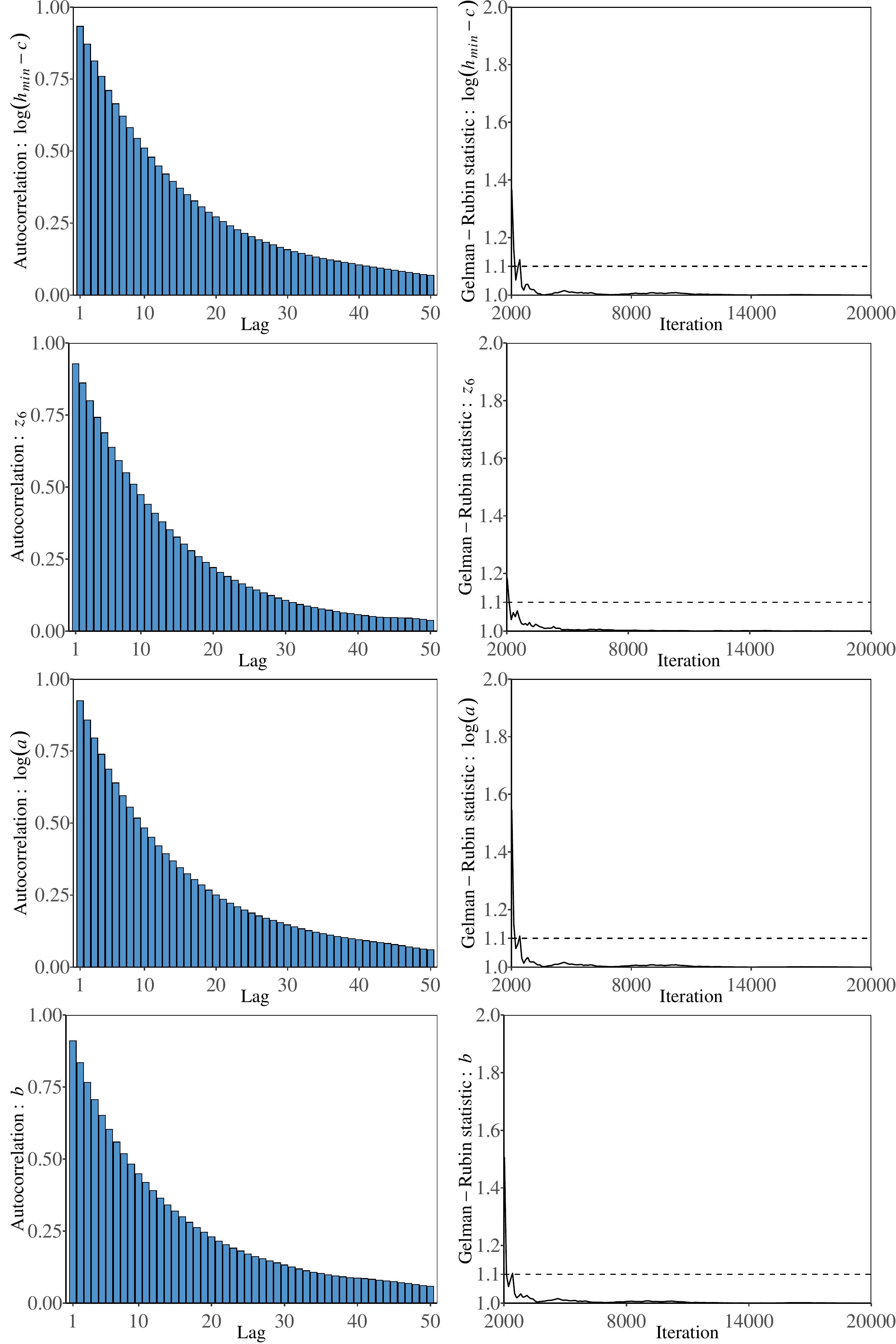}
\end{minipage}
\caption{Convergence diagnostics for Model 0 using the Jokulsa a Fjollum River dataset. Autocorrelation function (left panel) and the Gelman--Rubin statistic as a function of number of iterations (right panel) for the following parameters;
$\psi_{1} = \log(h_{\min}-c)$; $\psi_{10} = z_6$; 
$x_1=\log(a)$; $x_2=b$.}\label{acgrm1}
\end{figure}

\begin{figure}[hbt!]\centering
\begin{minipage}{0.95\linewidth}
\includegraphics[width=\linewidth]{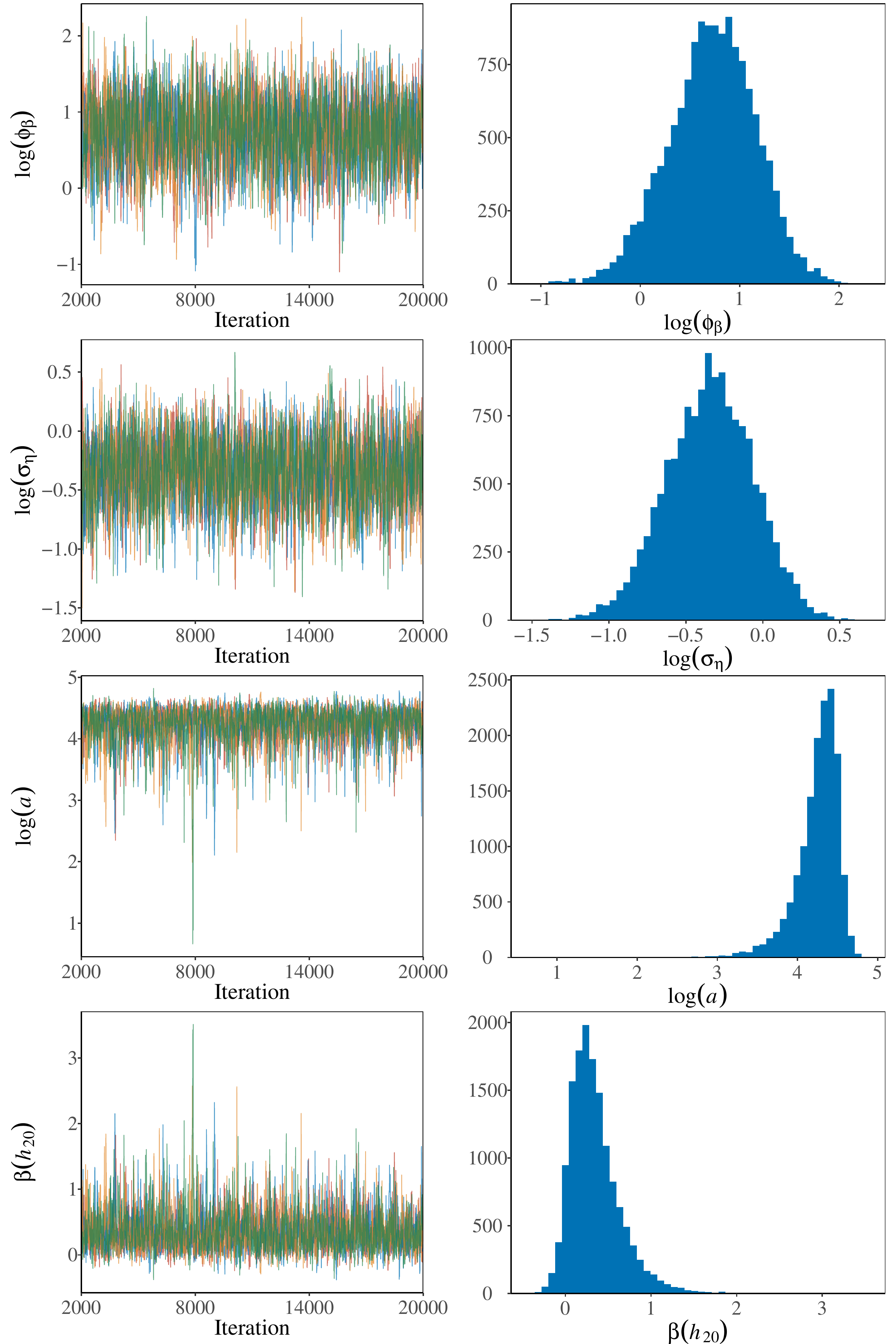}
\end{minipage}
\caption{Convergence diagnostics for Model 1 using the Jokulsa a Dal River dataset. Trace plots (left panel) of the four chains of simulation, and histograms (right panel) of samples from simulation for the following parameters;
$\psi_{3} = \log(\phi_\beta)$; $\psi_{4} = \log(\sigma_\eta)$; 
$x_1=\log(a)$; $x_{21}=\beta(h_{20})$.}\label{trhim2}
\end{figure}

\begin{figure}[hbt!]\centering
\begin{minipage}{0.95\linewidth}
\includegraphics[width=\linewidth]{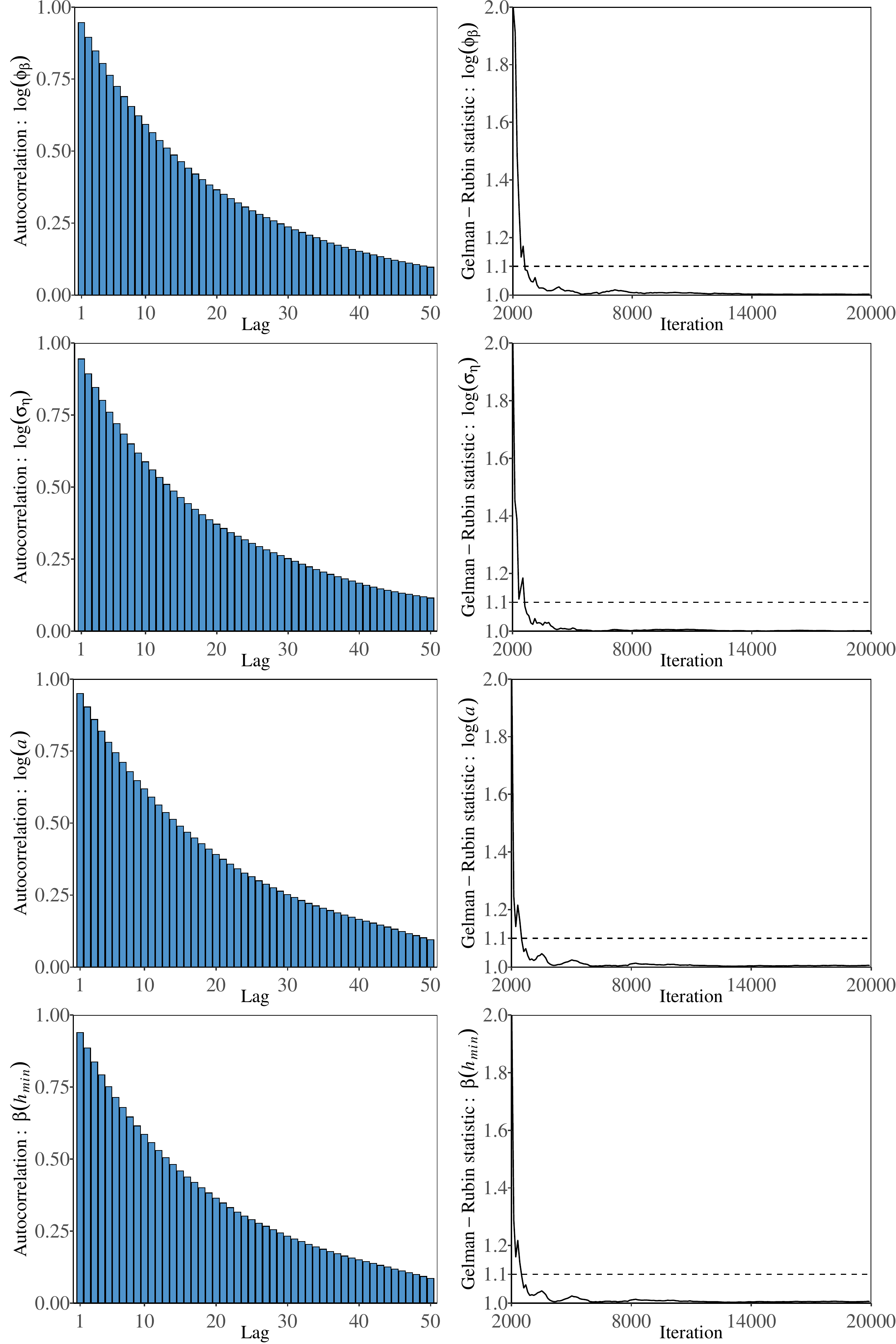}
\end{minipage}
\caption{Convergence diagnostics for Model 1 using the Jokulsa a Dal River dataset. Autocorrelation function (left panel) and the Gelman--Rubin statistic as a function of number of iterations (right panel) for the following parameters;
$\psi_{3} = \log(\phi_\beta)$; $\psi_{4} = \log(\sigma_\eta)$; 
$x_1=\log(a)$; $x_{22}=\beta(h_{20})$.}\label{acgrm2}
\end{figure}

Trace plots and histograms of four parameters in Model 0 for the Jokulsa a Fjollum River are shown in Figure \ref{trhim1} and Figure \ref{acgrm1} shows the autocorrelation function and the Gelman--Rubin statistic for the same parameters when using the same dataset.
Figure \ref{trhim2} shows the trace plots and histograms of four parameters in Model 1 for the Jokulsa a Dal River. Figure \ref{acgrm2} shows the autocorrelation function and the Gelman--Rubin statistic for the same parameters and the same dataset.
Results for other parameters in Model 0 and Model 1, and results based on the Nordura River dataset and the Skjalfandi River dataset are not shown. However, the convergence diagnostics plots for these parameters were the similar those shown in Figures \ref{trhim1}, \ref{acgrm1}, \ref{trhim2} and \ref{acgrm2}.
 
The Gelman--Rubin statistic plots show the rate of convergence with the value 1.1 as a reference point. Values above 1.1 indicate that sufficient convergence has not been reached. Trace plots and the autocorrelation function indicate the quality of simulated chains. A slowly decaying autocorrelation function indicates worse quality. Trace plots show how well the chains mix. A bad mixing of chains indicates worse quality. An example of a bad mixing is separate chains staying in separate regions over a large number of consecutive iterations and never crossing. Histograms give the marginal distributions under the assumption that convergence has been reached.
Figures \ref{trhim1}, \ref{acgrm1}, \ref{trhim2} and \ref{acgrm2} show that convergence according to the Gelman--Rubin statistic was reached after 10000 iterations (after burn-in), the autocorrelation between samples $50$ iterations apart (no thinning applied) was \textcolor{black}{less than} $0.2$, and the chains \textcolor{black}{of both Model 0 and Model 1} mix well.

